\documentclass[twocolumn]{aastex631}

\usepackage{bm}
\usepackage{booktabs}
\usepackage{tabularx}
\usepackage{commath}
\usepackage{xcolor}
\usepackage{amsmath}
\usepackage{graphicx}
\usepackage{float}

\newcolumntype{C}{>{\centering\arraybackslash}X}

\begin{document}

\title{Tracing High-$z$ Galaxies in X-rays with JWST and {\em Chandra}}

\author[0009-0000-3672-0198]{Aidan J. Kaminsky}
\affiliation{Department of Physics, University of Miami, Coral Gables, FL 33124, USA}
\author[0000-0002-1697-186X]{Nico Cappelluti}
\affiliation{Department of Physics, University of Miami, Coral Gables, FL 33124, USA}
\author[0000-0002-0797-0646]{Günther Hasinger}
\affiliation{TU Dresden, Institute of Nuclear and Particle Physics, 01062 Dresden, Germany}
\affiliation{DESY, Notkestraße 85, 22607 Hamburg, Germany}
\affiliation{Deutsches Zentrum für Astrophysik, Postplatz 1, 02826 Görlitz, Germany}
\author[0000-0003-2196-3298]{Alessandro Peca}
\affiliation{Eureka Scientific, 2452 Delmer Street, Suite 100, Oakland, CA 94602-3017, USA}
\affiliation{Department of Physics, Yale University, P.O. Box 208120, New Haven, CT 06520, USA}
\author[0000-0002-0930-6466]{Caitlin M. Casey}
\affiliation{The University of Texas at Austin, 2515 Speedway Blvd Stop
C1400, Austin, TX 78712, USA}
\affiliation{Department of Physics University of California Santa Barbara, CA,
93106, CA}
\affiliation{Cosmic Dawn Center (DAWN), Denmark}
\author[0000-0003-4761-2197]{Nicole E. Drakos}
\affiliation{Department of Physics and Astronomy, University of Hawaii, Hilo, 200 W. Kawili St., Hilo, HI 96720, USA}
\author[0000-0002-9382-9832]{Andreas Faisst}
\affiliation{Caltech/IPAC, MS 314-6, 1200 E. California Blvd. Pasadena, CA 91125, USA}
\author[0000-0002-0236-919X]{Ghassem Gozaliasl}
\affiliation{Department of Computer Science, Aalto University, PO Box 15400,
Espoo, FI-00 076, Finland}
\affiliation{Department of Physics, Faculty of Science, University of Helsinki,
00014-Helsinki, Finland}
\author[0000-0002-7303-4397]{Olivier Ilbert}
\affiliation{Aix Marseille Univ, CNRS, CNES, LAM, Marseille, France}
\author[0000-0001-9187-3605]{Jeyhan S. Kartaltepe}
\affiliation{Laboratory for Multiwavelength Astrophysics, School of Physics
and Astronomy, Rochester Institute of Technology, 84 Lomb Memorial Drive, Rochester, NY 14623, USA}
\author[0000-0003-2156-078X]{Alexander Kashlinsky}
\affiliation{Code 665, Observational Cosmology Lab, NASA Goddard Space Flight Center, Greenbelt, MD 20771, USA}
\affiliation{Dept of Astronomy, University of Maryland, College Park, MD 20742, USA}
\affiliation{Center for Research and Exploration in Space Science and Technology, NASA/GSFC, Greenbelt, MD 20771, USA}
\author[0000-0002-6610-2048]{Anton M. Koekemoer}
\affiliation{Space Telescope Science Institute, 3700 San Martin Drive, Baltimore, MD 21218, USA}
\author[0000-0002-9489-7765]{Henry Joy McCracken}
\affiliation{Institut d’Astrophysique de Paris, UMR 7095, CNRS, and Sorbonne Université, 98 bis boulevard Arago, F-75014 Paris, France}
\author[0000-0002-4485-8549]{Jason Rhodes}
\affiliation{Jet Propulsion Laboratory, California Institute of Technology, 4800 Oak Grove Drive, Pasadena, CA 91001, USA}
\author[0000-0002-4271-0364]{Brant E. Robertson}
\affiliation{Department of Astronomy and Astrophysics, University of California, Santa Cruz, 1156 High Street, Santa Cruz, CA 95064, USA}
\author[0000-0002-7087-0701]{Marko Shuntov}
\affiliation{Cosmic Dawn Center (DAWN), Denmark}
\affiliation{Niels Bohr Institute, University of Copenhagen, Jagtvej 128, 2200
Copenhagen, Denmark}
\author[0000-0002-2064-6429]{Joseph Sterling}
\affiliation{Department of Physics, University of Miami, Coral Gables, FL 33124, USA}



\begin{abstract}
We leverage \textit{JWST} data from the COSMOS-Web Survey in order to provide updated measurements on the auto-power spectrum of the now resolved Cosmic Infrared Background (CIB) and its coherence with the unresolved soft Cosmic X-ray Background (CXB) observed by \textit{Chandra} at $z > 6$. Maps of the CIB in the F277W and F444W NIRCam filters are constructed with sources fainter than $m_{AB} = 25$ and cross-correlated with the CXB in the [0.5-2] keV band. We find that on scales between $1$ and $1000''$ the CIB-CXB cross-power in both NIRCam filters is statistically significant with signal-to-noise ratios ($S/N$) of 4.80 and 6.20 respectively from redshifts $0\le z \le 13$. In our high-$z$ ($6\le z \le13$) interval we find coherence in both filters with a $S/N$ of 7.32 and 5.39 respectively. These results suggest that there are X-ray emitting galaxies resolved by \textit{JWST}, including star-forming galaxies (SFGs) and active galactic nuclei (AGNs). We fit the large-scale biasing of the IR sources producing the CIB as a function of $z$ with results consistent with prior measurements and place constraints on the CXB flux and biasing at low- and high-$z$. The CXB flux measurements presented in this study suggest that approximately 94\% of the [0.5-2] keV CXB is resolved, and this value is consistent within 2$\sigma$ with the complete resolution of the [0.5-2] keV CXB.
\end{abstract}

\section{Introduction} \label{sec:intro}
The Cosmic Near-Infrared Background (CIB) is the integrated radiation from star formation and other astrophysical processes, such as supermassive black hole (SMBH) accretion, spanning all redshifts. While most of the CIB has been resolved, a component remains from unresolved sources which cluster at large angular scales \citep{Kashlinsky2005,Kashlinsky2018}. This puzzling excess in the unresolved \textit{Spitzer} and \textit{AKARI} CIB fluctuations with respect to extrapolations from known galaxy populations have been reported by several groups \citep{Kashlinsky2005,Kashlinsky2012,Cooray2012,matsumoto}. The nature of this excess has been investigated \citep{helgason12,Helgason2014} and two main solutions have been proposed: (1) a high-$z$ signal originating from Population III stars in primordial galaxies  \citep{Kashlinsky2005,Kashlinsky2025} and (2) a z $\sim$ 1--4 signal from Intra-Halo Light (IHL), the result of stripped material (gas and stars) from galaxy mergers \citep{Cooray2012}.

Uncovering the origins behind this excess in the CIB requires knowing the properties of faint galaxies detected in the IR, such as their redshifts, star formation histories, and black hole activity. In the age of the \textit{James Webb Space Telescope} (\textit{JWST}), it is possible to resolve the CIB above 1 $\mu$m into discrete sources down to $m_{AB}\sim$ 30, allowing for a better understanding of the populations imprinted on the faint CIB fluctuations.

On the other hand, the Cosmic X-ray Background (CXB) \citep{giacconi} is the result of the cumulative X-ray emission dominated by accretion onto SMBHs with a small fraction attributed to normal galaxies. X-ray surveys with ROSAT, {\em Chandra} and XMM-{\em Newton} resolved $>$90\% of it into discrete point-sources \citep{hm06,moretti, Cappelluti2017cxb}. The nature of the unresolved $<$10\% CXB and its exact recipe is still debated \citep{cap12,Helgason2014} in particular when it comes to determining the contribution of high-$z$ (z$>$6) SMBH accretion. Understanding this is critical, a measurement of such emission at high-$z$ can inform us on the nature of SMBH seeds and the potential contributions of accretion on the reionization of the Universe.

Interestingly, it has been discovered that the unresolved CIB anisotropies show a high level of coherence at large angular scales ($\sim 1000''$) with those of the unresolved \textit{Chandra} [0.5-2] keV CXB \citep{Cappelluti2013,Cappelluti2017,mw16,Li2018}. 
The origin of the CIB-CXB coherence has been a matter of debate and several explanations have been proposed. These include X-ray binary emission in star-forming galaxies (SFGs) and obscured active galactic nuclei (AGNs), both populations being unresolved by \textit{Spitzer} and \textit{Chandra}. Specifically in the high-$z$ regime, primordial black holes (PBHs) \citep{Kashlinsky2016,Ricarte2019,Hasinger2020,Cappelluti2022,Kashlinsky2025} and direct collapse black holes (DCBHs) \citep{Yue2013,Ricarte2019} are proposed to have observable contributions to the CIB-CXB cross-power spectrum, specifically at large angular scales given that DCBHs would be highly biased with respect to the underlying dark matter distribution.

Using angular Fourier analysis has proved thus far to be a powerful tool for studying the CIB and its coherence with the soft CXB. This technique can provide constraints on the number density of sources producing these unresolved backgrounds and how they are spatially distributed. In particular, statistically significant measurements of the CIB auto- and CIB-CXB cross- power signals as a function of redshift hold important information regarding how SFGs and AGNs evolve in their large-scale structures \citep{Powell2020,Paquereau2025}.

The first two years of \textit{JWST} operations showed that at z $>$ 5 a larger than expected population of overmassive SMBHs live in primordial galaxies that seem to be more massive than expected { \citep{Pacucci2023,Natarajan2024}}. However, as of today, there have been few serendipitous detections of X-rays from galaxies at $z>6$ \citep{2024NatAs...8..126B}. This hinders our ability to place constraints on the first AGNs that populated the Universe in the first hundreds of millions of years after the Big Bang. { However, this black hole population may have a measurable contribution to the CIB-CXB cross-power signal due to being more luminous and highly biased given that they likely occupy larger dark matter halos. A Fourier analysis of the CIB and CXB anisotropies would allow us to investigate these properties of such overmassive SMBHs.}    

In this paper, we leverage the exquisite wide-field photometric survey of the COSMOS-Web field \citep{Casey2023} to study the coherence between the resolved CIB and the unresolved \textit{Chandra} COSMOS-Legacy CXB fluctuations. With the sources contributing to the resolved CIB, we can recreate the diffuse CIB as observed with \textit{Spitzer} surveys by placing sources on the COSMOS-Web field according to their positions as if they
were unresolved. This paper is organized as follows. We describe the data employed and outline the computations of the auto- and cross-power spectra in \S 2. In \S 3, we present our analysis of the angular power spectra followed by a discussion \S 4 where we model the computed angular power spectra. Finally, we summarize our results in \S 5.

\section{Dataset and Methodology}
\subsection{Chandra}
We use CXB fluctuation maps in the [0.5-2] keV band constructed and analyzed in \cite{Li2018}. This data was taken as part of the COSMOS-Legacy survey with a 1.5 deg$^2$ area and total observing time of 4.6 Ms \citep{Civano2016}. Observations are recorded by arrival time and placed into A and B images, where the first image is for even events and the second is for odd ones. From these images, a mosaic signal map ($A+B$) and noise map ($A-B$) are created. Because the maps have the same exposure time, the only difference in the $A$ and $B$ images will be instrumental effects contributing to the noise of our computations.

Additionally, resolved X-ray sources detected in the field \citep{Civano2016} are masked using circular masks with radii of 7$''$. This method removes over 90$\%$ of the source brightness when factoring off-axis PSF degradation and mosaicing, with the remaining flux having no significant contribution to the cross-power spectra computed in \cite{Li2018}. Furthermore, extended emission was masked using the maps of \cite{Finoguenov2007}. For further information regarding the data reduction and masking, see \cite{Li2018} as maps were directly provided by the authors.
\subsection{COSMOS-Web CIB Map-Making}

In this study, we use the \texttt{SourcExtractor++} \citep[SE++,][]{Bertin2020} photometric catalog of sources in the COSMOS-Web field, a contiguous 0.54 deg$^2$  NIRCam imaging survey \citep{Casey2023}. Imaging is done in both the NIRCam filters F115W, F150W, F277W, and F444W and the MIRI filter F770W (although the latter is only done in 0.19 deg$^2$). The catalog is developed by extracting photometry using \texttt{SE++}: a $\chi^2$-image is created by co-adding the four maps (for each NIRCam filter) and the source detection is performed on the image. Detections are then modeled with a double Sérsic profile and the best-fit model is PSF-matched with the NIRCam and MIRI maps \citep{Sersic1968}. From the PSF-matching, photometric data for each source in the different maps are obtained. The exact procedures used in constructing the COSMOS-Web catalog(s) can be found in Shuntov et al. (in prep). We simulate CIB maps in the F277W and F444W filters and focus our analysis on the latter.

\begin{figure} 
    \centering
    \includegraphics[width=\columnwidth]{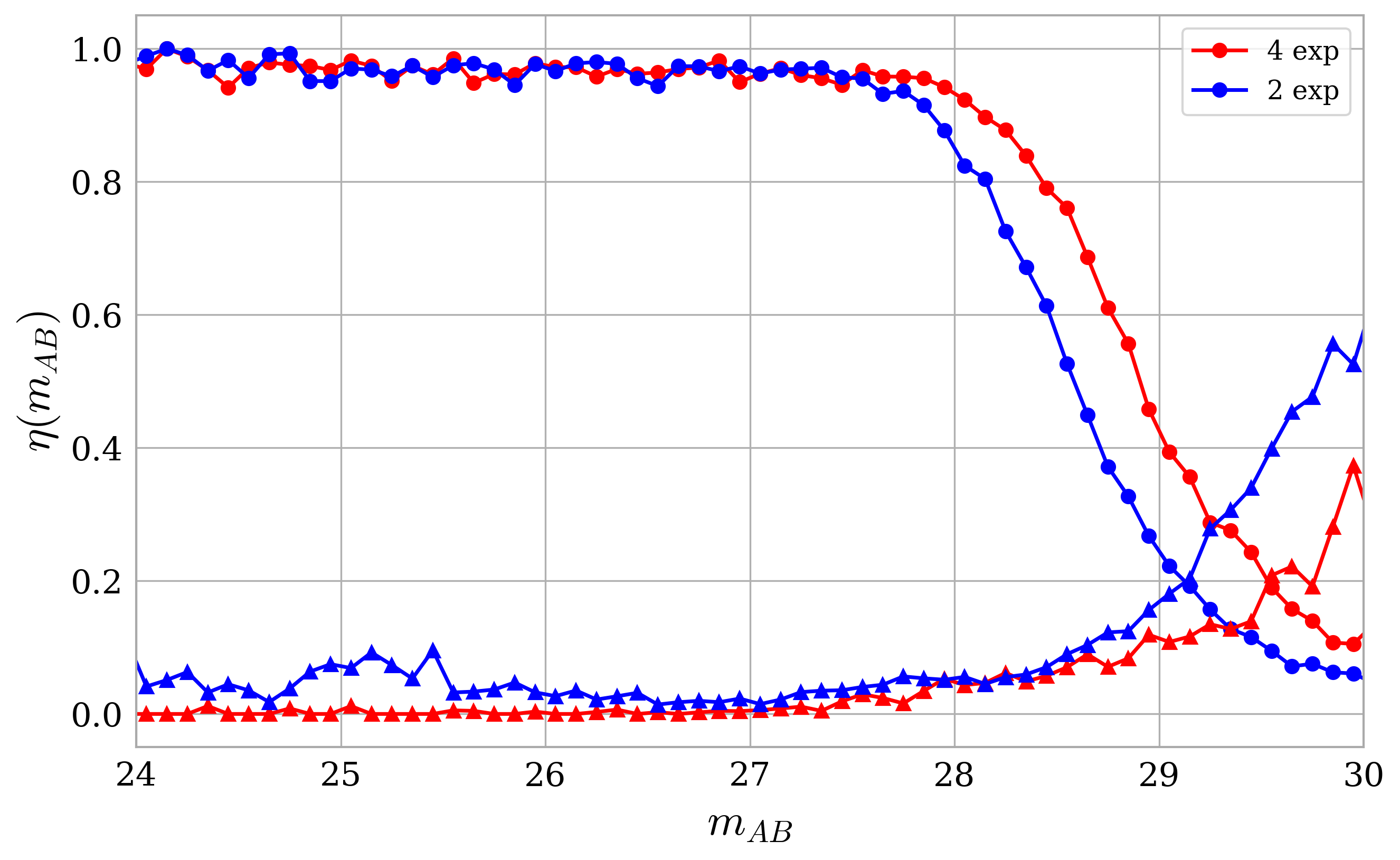} 
    \includegraphics[width=\columnwidth]{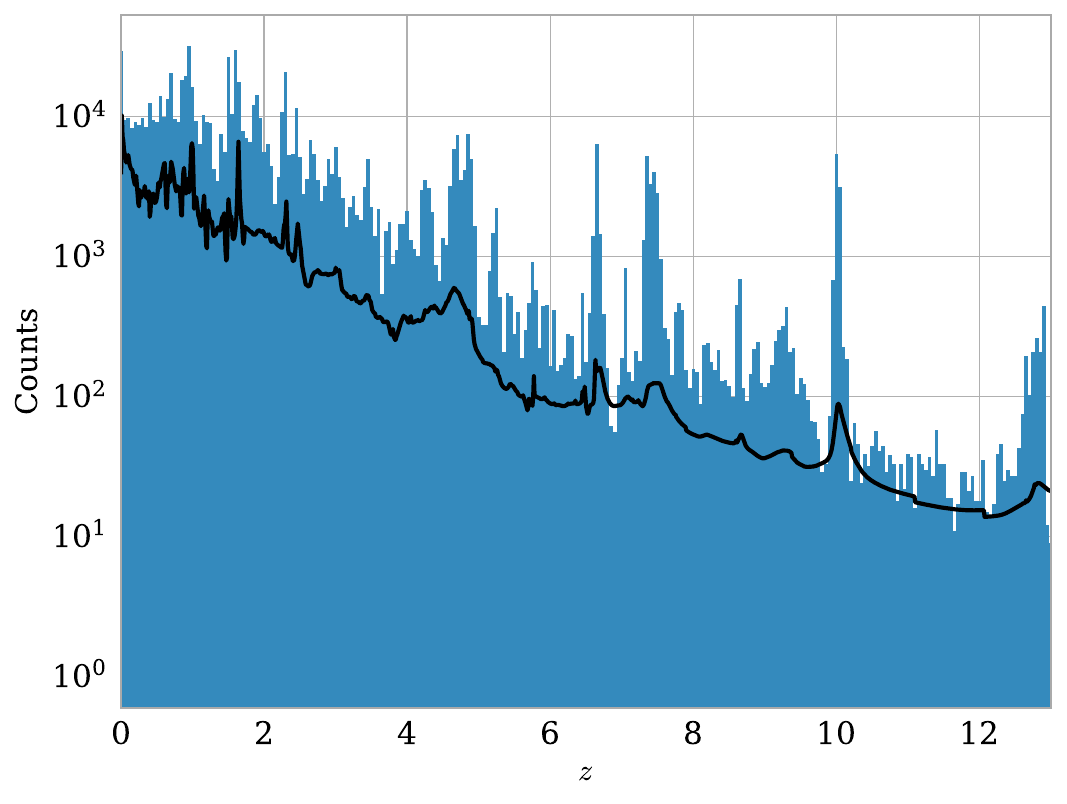} 

  \caption{Top: The selection function obtained with the deeper PRIMER survey in the COSMOS field. The different colors signify the completeness for regions of COSMOS-Web with either 2 (blue) or 4 (red) exposures. The circles correspond to the completeness and the triangles to the contamination. Bottom: Best-fit redshift distribution of sources in the COSMOS-Web catalog (blue) and overplotted is the summed z-PDF for all sources (black).} 
  \label{fig:CW_distributions}
\end{figure}

\subsubsection{CIB Map Production}
Sources with [F277W] \& [F444W] $> 25$ are placed on the sky according to their positions as point sources. The lower threshold of $m_{AB} = 25$ was chosen so that our CIB maps are only populated by discrete sources which were unresolved by \textit{Spitzer} \citep[see e.g.][see \S \ref{subsec:auto-power}]{Kashlinsky2012}. We also place an upper $m_{AB}$ cutoff of 29 which mitigates errors due to contamination (see Fig. \ref{fig:CW_distributions}). The flux density (typically $f_{\nu}$, but we will use $F$) is computed for each ``point source" assuming a surface area of 0.98$^{\prime\prime}$ $\times$ 0.98$^{\prime\prime}$, the same pixel scale as in the CXB fluctuation maps. To account for incompleteness, the flux of each source was weighted by its corresponding selection function $\eta(m_{AB})$. 

The selection function (or completeness) of the \texttt{SE++} catalog is computed using PRIMER (GO \#1837), a much deeper survey in the COSMOS and UDS fields (\cite{Casey2023}; see Shuntov et al. (in prep.) for further details). Direct measurements of COSMOS-Web sources are compared to those in PRIMER \citep{2021jwst.prop.1837D}, allowing for a robust assessment of the completeness as a function of magnitude. As discussed in more detail in \cite{Casey2023}, COSMOS-Web has a variable depth that changes with the number of exposures $N$. Of the 0.54 deg$^2$ area, $\sim 50.8\%$ of it has $N=2$ and $\sim 47.0\%$ has $N=4$ (see Fig. \ref{fig:CW_distributions}). Because of the non-uniform sensitivity of the field, a striping effect becomes evident which introduces artificial fluctuations in the subsequent auto- and cross- power spectra. More specifically, an artificial spike in the CIB fluctuations occurs at $\sim 200''$. This effect is mitigated by flattening the $N=4$ sources according to the $N=2$ selection function. For the sources with $N=2$, sources are separated into $m_{AB}$ bins $\Delta m_{AB}$ where the number of sources in each bin is $n_{N=2}$, which is multiplied by the ratio of the coverage areas ($A_{N=4}/A_{N=2}$). Once this term $n_{N=2}(A_{N=4}/A_{N=2})$ is computed, sources with $N=4$ are randomly removed until $n_{N=4}=n_{N=2}(A_{N=4}/A_{N=2})$. 1000 iterations of this procedure are performed with negligible differences. Once this procedure is applied, all sources are weighted with the $N=2$ selection function. Prior to flat fielding, the mean flux of the $N=2$ and $N=4$ sources was
$6.1978\times10^{-11}$ nW/m$^2$ and $5.4566\times10^{-11}$
nW m$^{-2}$ respectively. When applying the flat-fielding technique, the $N=4$ mean flux becomes $6.1955\times10^{-11}$
nW m$^{-2}$ and the subsequent power spectra is without the artificial peaks introduced by the striping.

After correcting each source for incompleteness, three maps are created by dividing the catalog into three photometric redshift (photo-z) bins $\Delta z$ = [0 -- 3], [3 -- 6],  and [6 -- 13] (see Fig. \ref{fig:CW_distributions}). We do this following the outlined procedure in \cite{Allevato2016} and \cite{Powell2020}. The photometric redshift probability distribution function (z-PDF) of each source is computed using the \texttt{Le Phare} software \citep{Ilbert2006,Arnouts1999}. For more detailed information regarding the fitting method, see Shuntov et al. (in prep). The flux of each source is weighted by the z-PDF in each redshift bin and then placed in the respective CIB map. That is, the flux of each source in a given bin is $F(\Delta z) = \int_{z_1}^{z_2}p(z)dz \cdot~F$ where the subscripts 1 and 2 are for the lower and upper bounds of each bin respectively. Essentially, the flux is ``spread'' into each of the different redshift bins. We also analyzed a combined CIB map made of sources from z = 0 and z = 13 using the same flux-weighting method. In total, there are eight different artificial CIB maps created for each $\Delta z$ and NIRCam filter ([0 -- 3, 3 -- 6, 6 -- 13, 0 -- 13] $\times$ [F277W, F444W]).

Furthermore, sources flagged in the COSMOS-Web catalog are masked. This includes sources that are present in the \textit{JWST} and \textit{Hyper Suprime-Cam} (or \textit{HSC}) star masks, in addition to sources that have \textit{Chandra} counterparts. Additionally, sources with contaminated photo-z measurements (Shuntov et al. in prep) are removed. Finally, all masked sources are combined in one map, which is joined with the \textit{Chandra} mask \citep{Li2018}, resulting in one singular mask applied to all CIB and CXB maps.

\subsection{Fourier Analysis of the CIB and CXB Fluctuations}

The CIB and CXB flux maps are then converted into fluctuation maps with: \( \displaystyle\delta F(\mathbf{x}) = F(\mathbf{x}) - \langle F \rangle\), where $F(\mathbf{x})$ is the flux at a given pixel and $\langle F \rangle$ is the average flux. The Fourier transform of these fluctuations, \( \displaystyle \Delta (\mathbf{q}) = \int \delta Fe^{-i\mathbf{x}\cdot \mathbf{q}}d^2x \), is used to compute the angular auto-power spectrum \( \displaystyle P(\mathbf{q}) = \langle {| \Delta (\mathbf{q}) |}^2 \rangle\). 
The goal of our study is to analyze the cross-power spectrum and coherence of the CIB and CXB up to $\sim$0.5 deg scale. To avoid masking effects on the auto- and cross-power spectra we require a significant fraction of unmasked pixels \citep{Kashlinsky2005,Kashlinsky2012}. Here we obtained an unmasked fraction of $\sim$65\% hence Fourier analysis can be used instead of two-point statistics \citep{Kashlinsky2005}.

\begin{figure*}[ht!]
\begin{center}
    \includegraphics[width = \textwidth]{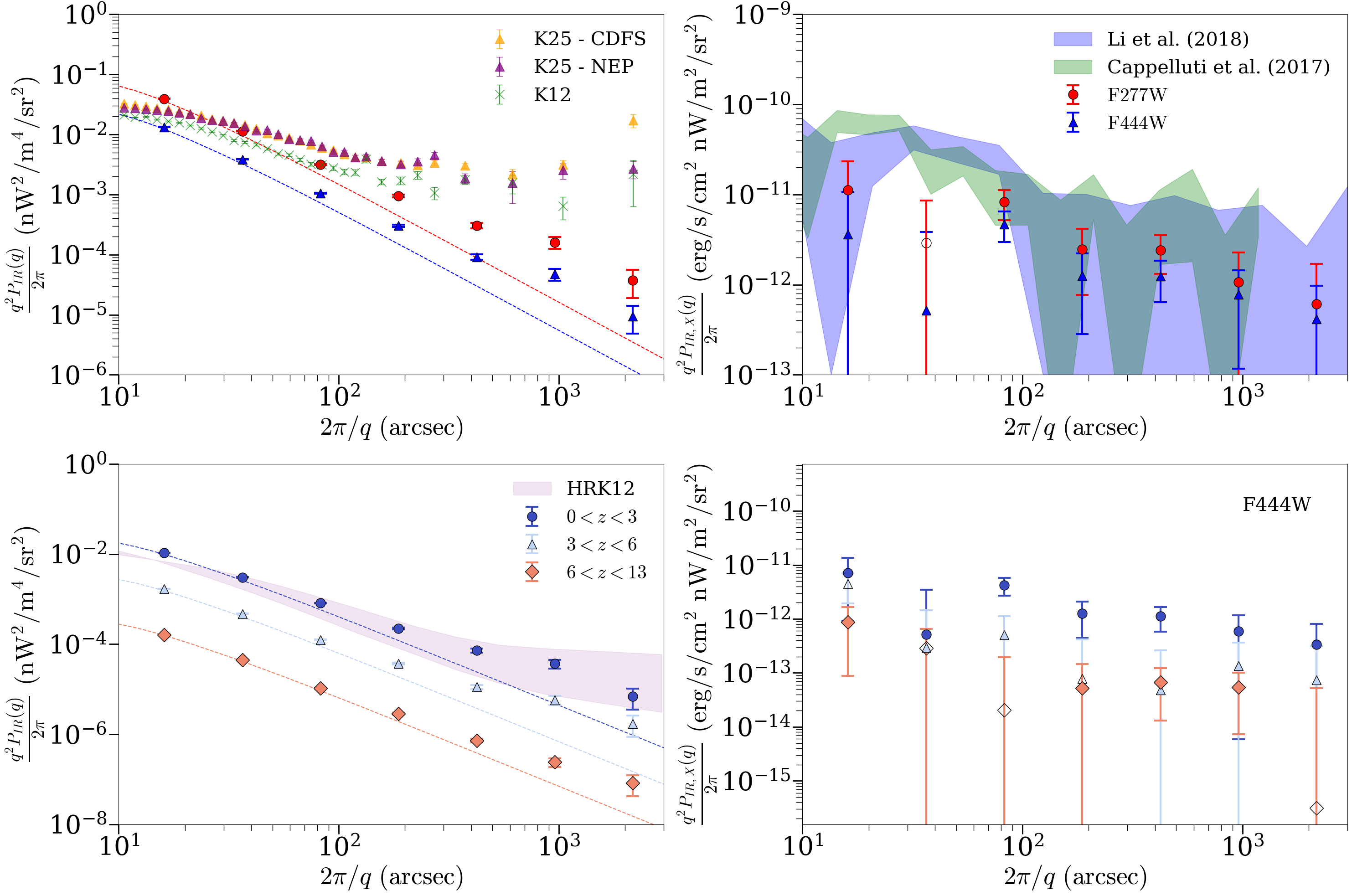}
\end{center}
\caption{{ \textit{Upper Left}: Auto-Power spectra of the CIB fluctuations for F277W and F444W CIB fluctuation maps, denoted with the red circles and blue triangles respectively. The dashed lines indicate the shot noise levels computed directly from the COSMOS-Web source counts. The auto-power spectra and the associated shot noise lines are convolved with the \textit{Spitzer} beam. The  orange and purple triangles correspond to the same computation done in \cite{Kashlinsky2025} (4.5 $\mu$m, $m_{lim} \sim 24.5$) in the CDFS and NEP fields, respectively. The green crosses correspond to the CIB auto-power spectrum measured in in \cite{Kashlinsky2012} (4.5 $\mu$m, $m_{lim} \sim 25$). \textit{Upper Right}: Cross-Power spectra of the F277W and F444W and the [0.5-2] keV CXB. The filled-in green and blue regions correspond to the same computations done in \cite{Cappelluti2017} and \cite{Li2018}, respectively. \textit{Bottom Left}: Auto-Power spectra for each redshift bin of the F444W map, with the circles, triangles, and diamonds corresponding to the $\Delta z = [0-3], [3-6], [6-13]$ bins respectively. The dashed lines are used again to indicate the shot noise levels computed directly from the source counts. Again, the auto-power spectra and corresponding shot noise lines are convolved with the \textit{Spitzer} beam. Overplotted in purple is the auto-power spectrum of known populations at $z \leq 6$ \citep{helgason12}. \textit{Bottom Right}: Cross-Power spectra for each redshift bin of the F444W map. Errors are reported at the $1\sigma$ confidence level.}}
\label{fig:all_auto_cross}
\end{figure*}

Following \cite{Cappelluti2013,Cappelluti2017,Li2018}, the auto-power of the $A-B$ map is subtracted from the auto-power of the $A+B$ map to retrieve the clean auto-power spectrum of the CXB. Furthermore, we compute the cross-power spectra of the CIB and CXB expressed as \( \displaystyle P_{IR,X}(\mathbf{q}) = \langle {\Delta_{IR} (\mathbf{q})\Delta_{X}^* (\mathbf{q})}\rangle \). This is done by cross-correlating the CIB fluctuation maps with the X-ray $\delta F_{A+B}(\mathbf{x})$ and $\delta F_{A-B}(\mathbf{x})$ maps separately and then taking the difference to construct the clean CIB-CXB cross-power spectrum. In total, there are eight CIB auto-power spectra (for [0 -- 3, 3 -- 6, 6 -- 13, 0 -- 13] $\times$ [F277W, F444W]) and eight corresponding cross-power spectra, all with the same mask applied.

For both the auto- and cross-power spectra we define the uncertainty at a given angular scale \( \displaystyle \frac{2\pi}{q} \) using Poissonian estimators with \( \displaystyle \frac{P_{IR}}{\sqrt{0.5N_q}}\) and \( \displaystyle \sqrt{\frac{P_{IR}P_X}{N_q}} \) for the auto- and cross-power spectra respectively. In the previous expressions, $0.5N_q$ is defined as the number of independent Fourier elements at a given $q$. In this work we also quantify the broadband auto- and cross- power signals defined as 
\begin{equation}
\langle P_{IR}\rangle = \frac{\sum \sigma_{IR}(q) \cdot P_{IR}(q)}{\sum \sigma_{IR}(q) }
\end{equation}
and
\begin{equation}
\langle P_{IR,X}\rangle = \frac{\sum \sigma_{IR,X}(q) \cdot P_{IR,X}(q)}{\sum \sigma_{IR,X}(q) }
\end{equation}
where $\sigma_{IR}$ and $\sigma_{IR,X}$ are the uncertainties in the auto- and cross-power measurements.

The relative contribution of the emission of one background to the other can be parameterized by a coherence term $\mathcal{C}$, expressed as 
\begin{equation}
\mathcal{C}(q) = \frac{P_{IR,X}^2}{P_X P_{IR}}
\label{eq:coherence1}
\end{equation}
This quantity indicates how much of one signal (either the CXB or CIB auto-power) is present in the other. Looking at the two extremes, $\mathcal{C} = 0$ points to the scenario that there is no contribution from one background to the other, and alternatively $\mathcal{C} =1$ means that all of the signal from one background contributes to the other.

\section{Results}

\subsection{CIB Auto-Power Spectra}\label{subsec:auto-power}
In Fig. \ref{fig:all_auto_cross} we present the CIB auto-power spectra computed in the broad ([0 -- 13]) and narrow ([0 -- 3, 3 -- 6, 6 -- 13]) redshift bins. In the broad $\Delta z$ bin, upon initial inspection a flat (shot) noise component at low angular scales and an increasing (clustering) component at larger angular scales is visible for both the F277W and F444W maps. The shot noise is estimated as the auto-power signal measured at $\sim 1''$ given a $\Delta \log(q)$ of $0.35$, yielding a value of $6.076$ $\pm$ $0.006$ $\times$ $10^{-11}$ and $2.067$ $\pm$ $0.002$ $\times$ $10^{-11}$ nW$^2$ m$^{-4}$ sr$^{-1}$, for the F277W and F444W bands respectively. We can compare these values to calculations of the shot noise using the COSMOS-Web source catalogs with the following expression 
\begin{equation}
P_{IR}^{\textrm{SN}}=\int_{m_{lim}}^{\infty}f^2(m_{AB})\frac{dN}{dm_{AB}}dm_{AB}
\label{eq:aps_sn}
\end{equation}
where $f(m_{AB})$ is the flux and ${dN}/{dm_{AB}}$ is the differential number counts per deg$^2$. Calculating this term in the $\Delta z = [0-13]$ bin results in a value of $6.072 \times 10^{-11}$ and $2.064 \times 10^{-11}$ nW$^2$ m$^{-4}$ sr$^{-1}$ for the F277W and F444W bands respectively, consistent with the directly fitted values. This consistency serves as a sanity check for the CIB map creation technique. 

The signal-to-noise ratio $S/N$ of the CIB auto-power spectrum in the broad redshift bin for the both the F277W and F444W maps is $\sim 11$ when sampling angular scales from $\sim 1-3000''$. Furthermore, we compute the average auto-power signal $\langle P_{IR} \rangle$ at scales above $100''$ resulting in measurements of $6.01$ $\times$ $10^{-10}$ and $1.57$ $\times$ $10^{-10}$ nW$^2$ m$^{-4}$ sr$^{-1}$ for the F277W and F444W maps respectively. Both values are found at a $S/N$ of $\sim 5$, showing a statistically significant signal due to the large-scale spatial distribution of galaxies (both in the linear and non-linear regimes).

{ Furthermore, we can compare the auto-power spectrum in the F444W filter with the measurements of the unresolved CIB analyzed in \cite{Kashlinsky2012} and \cite{Kashlinsky2025}. In the former analysis, auto-power spectra were computed from the Ultra-Deep Survey and Extended Groth Strip (UDS and EDS respectively; \citep{Fazio2011}) to a shot noise level corresponding to $m_{lim}\approx 25$. The UDS and EDS fields are $\sim 0.12$ and $\sim 0.14$ deg$^2$ respectively. In the latter work, observations are made of the North Ecliptic Pole (NEP) and Chandra Deep Field South (CDFS) fields (see Program IDs 13153 and 13058; \cite{Capak2016}). The FOV of the NEP and CDFS fields are 1.86 and 1.96 deg$^2$ respectively.} 

{ The auto-power spectra computed in both \cite{Kashlinsky2012} and \cite{Kashlinsky2025} are reasonably consistent with our results at small scales, with differences originating from the $m_{lim}$ that we adopt in this work in addition to cosmic variance in the fields used. More specifically, in \cite{Kashlinsky2012} the shot noise levels measured imply a $m_{lim}$ value of 24.5 -- 25, while in \cite{Kashlinsky2025} the $m_{lim}$ estimate is closer to $\sim 24.5$, explaining the small discrepancies at lower angular scales. However, the large-scale component detected in both \cite{Kashlinsky2012} and \cite{Kashlinsky2025} is significantly higher the the clustering component found here (see Fig. \ref{fig:all_auto_cross}). The primary reason for this discrepancy can be attributed to the fact that the large-scale signal from new populations found in \cite{Kashlinsky2012} and \cite{Kashlinsky2025} are characterized by faint magnitudes lower than 28 -- 30, pushing these sources out of the $m_{AB}$ range chosen for this work. This is supported by the lack of an observed co-evolution between the shot noise and large-scale power found in previous studies, suggesting a very faint/diffuse source population. To quantify this divergence at higher angular scales, we can evaluate $P_{IR}/P_{K12}$ and $P_{IR}/P_{K25}$ (where the subscripts denote 
\cite{Kashlinsky2012} and \cite{Kashlinsky2025} respectively) at $2\pi/q\approx1000''$ using the data points in Fig. \ref{fig:all_auto_cross}. We find that the relative fractions are $\sim 2\%$ and $\sim 8\%$, for the \cite{Kashlinsky2012} and \cite{Kashlinsky2025} power spectra respectively. Furthermore, we note that masking effects will not  warp the large-scale clustering as shown in \cite{Kashlinsky2012}.}

The CIB auto-power signals in the redshift bins $\Delta z$ = [0 -- 3], [3 -- 6], and [6 -- 13] for both the F277W and F444W maps are statistically significant at the $\sim 11\sigma$ level (sampling scales $\sim 1-3000''$). For the F277W maps, the directly estimated shot noise values are $4.859$ $\pm$ $0.005$ $\times$ $10^{-11}$, $7.992$ $\pm$ $0.007$ $\times$ $10^{-12}$, and $4.315$ $\pm$ $0.004$ $\times$ $10^{-13}$ nW$^2$ m$^{-4}$ sr$^{-1}$ for the $\Delta z$ = [0 -- 3], [3 -- 6], and [6 -- 13] bins respectively. These values can be compared to the shot noise power computed directly from the source counts, which is $4.864$ $\times$ $10^{-11}$, $7.964$ $\times$ $10^{-12}$, and $4.312$ $\times$ $10^{-13}$ nW$^2$ m$^{-4}$ sr$^{-1}$ for each respective $\Delta z$ bin. The estimated and directly computed shot noise values for each redshift interval are consistent within $1\sigma$ except for the intermediate bin $\Delta z = [3-6]$, which differs at the $\sim 4\sigma$ level. For each $\Delta z$ we compute the average auto-power signal $\langle P_{IR} \rangle$ at scales above $100''$, resulting in values of $4.16$ $\times$ $10^{-10}$, $9.24$ $\times$ $10^{-11}$, and $1.72$ $\times$ $10^{-12}$ nW$^2$ m$^{-4}$ sr$^{-1}$, each with a $S/N$ of 4.98, 5.2, and 5.0 respectively.

The previous analysis of the CIB observed in the F277W filter is extended to the F444W maps. The directly estimated shot noise values are $1.659$ $\pm$ $0.002$ $\times$ $10^{-11}$, $2.584$ $\pm$ $0.002$ $\times$ $10^{-12}$, and $2.634$ $\pm$ $0.002$ $\times$ $10^{-13}$ nW$^2$ m$^{-4}$ sr$^{-1}$ for each respective $\Delta z$ bin. The shot noise levels computed from the source catalogs for each redshift interval are $1.657$ $\times$ $10^{-11}$, $2.580$ $\times$ $10^{-12}$, and $2.639$ $\times$ $10^{-13}$ nW$^2$ m$^{-4}$ sr$^{-1}$, which is consistent with the direct estimates within $\sim1.5\sigma$. For the $\Delta z$ = [0 -- 3], [3 -- 6], and [6 -- 13] maps, the $\langle P_{IR} \rangle$ values above 100'' are $1.18$ $\times$ $10^{-10}$, $2.68$ $\times$ $10^{-11}$, and $1.26$ $\times$ $10^{-12}$ nW$^2$ m$^{-4}$ sr$^{-1}$ respectively. The $S/N$ of each value is 5.0, 5.2, and 5.2 respectively. It is important to note that when transitioning from the broad to narrow $\Delta z$ bins there is a systematic uncertainty introduced that is $\le 7\%$, which is found in both the number counts and maps generated (which are done independently). 

Furthermore, we can pay attention specifically to the [0 -- 3] and [3 -- 6] auto-power spectra, and compare our results to the reconstructed auto-power from known galaxies at $z < 6$ computed in \cite{helgason12}. In their work, they take 233 luminosity functions and fit their evolution with redshift to obtain number counts below the \textit{Spitzer} flux limit. For $m_{lim}=25$, the corresponding upper and lower limits on the CIB shot noise power is $\sim 0.6 $ and $\sim 3 $ $\times10^{-11}$ nW$^2$ m$^{-4}$ sr$^{-1}$ respectively. We thus find that our shot noise level of $1.93 \times 10^{-11}$ nW$^2$ m$^{-4}$ sr$^{-1}$, computed directly from the COSMOS-Web source counts, is consistent with their results \citep{helgason12}. Additionally, we find that the large-scale power computed in this study is within the range of allowed models from \cite{helgason12} (see Fig. \ref{fig:all_auto_cross}).

\subsection{CIB-CXB Cross-Power Spectra} \label{subsec:cross-power}
In this study we report cross-power spectra with $S/N$ measurements $>4\sigma$ at angular scales $1-1000''$. We find that the $S/N$ of the cross-power signals in the $\Delta z = [0-13]$ interval are $\sim 4.8$ and $6.2$, corresponding to the F277W and F444W maps, respectively. 

The $S/N$ of the narrower $\Delta z$ cross-power signals are shown in Table \ref{tab:avgcp} for both the F277W $\times$ [0.5-2] keV and F444W $\times$ [0.5-2] keV cross-correlations.

\begin{table}[h]
\centering
{Average Cross-Power Spectra}
\begin{tabularx}{\columnwidth}{l c c c c}
\toprule
\colhead{$\Delta z$} & \colhead{$\langle P_{IR,X} \rangle_{\textrm{F277W}}^a$} & \colhead{$S/N_{\textrm{F277W}}$} & \colhead{$\langle P_{IR,X} \rangle_{\textrm{F444W}}^a$} & \colhead{$S/N_{\textrm{F444W}}$} \\
\midrule
0 -- 3 & $25.2 \pm 5.81$ & $4.34$ & $17.2 \pm 3.18$ & $5.40$ \\
3 -- 6  & $3.95 \pm 0.24$ & $1.67$ & $3.73 \pm 1.26$ & $2.95$  \\
6 -- 13  & $2.48 \pm 0.34$ & $7.32$ & $1.47 \pm 0.27$ & $5.39$   \\
0 -- 13  & $23.8 \pm 5.77$ & $4.80$ & $22.5 \pm 3.61$ & $6.20$ \\ 
\bottomrule

\end{tabularx}
\tablenotetext{a}{$\times 10^{-19}$ erg/s/cm$^2$ nW/m$^2$/sr}

\caption{The average cross-power signal is computed in a each redshift bin ranging from the angular scales $1-1000''$. The top and bottom sections of the table are  measurements  using the F277W and F444W maps, respectively.}
\label{tab:avgcp}
\end{table}


\section{Discussion}\label{sec:discussion}
\begin{figure*}[ht!]
\begin{center}
    \includegraphics[width = \textwidth]{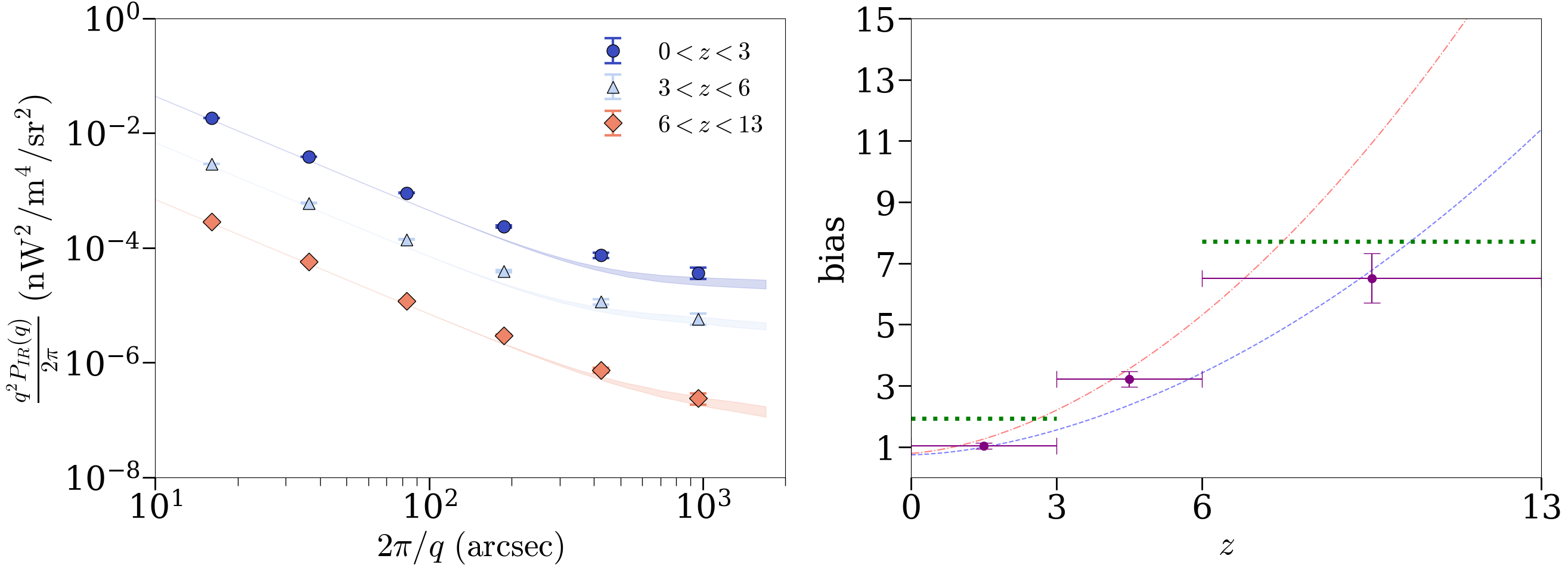}
\end{center}
\caption{\textit{Left}: Modeled F444W CIB auto-power spectra for each $\Delta z$ bin. The circles, triangles, and diamonds correspond to the $\Delta z = [0-3], [3-6], [6-13]$ bins respectively. \textit{Right}: Best-fit bias $\tilde{b}_{IR}$ as a function of $z$. {The dashed blue and dash-dotted red lines correspond to the  biasing of dark matter halos with masses of $10^{10}$ $M_{\odot}h^{-1}$ and $10^{11}$ $M_{\odot}h^{-1}$ respectively \citep{SMT2001}.} The green dotted lines correspond to the lower limits of the biasing of coherent X-ray sources.}
\label{fig:aps_bias_fits}
\end{figure*}
\subsection{Modeling the CIB Auto-Power Spectra}\label{subsec:cib}
Direct knowledge of the populations contributing to the CIB (i.e. source counts) allows us to constrain the large-scale clustering of these populations as a function of redshift. As mentioned previously, the CIB angular power can be written as
\begin{equation}
P_{IR}^{\textrm{tot}}(q)=P^{\textrm{SN}}_{IR}+P^{\textrm{Cl}}_{IR}(q)
\end{equation}
where the superscripts ``SN" and ``Cl" denote the shot noise and clustering components respectively. The shot noise component of the CIB auto-power spectra was computed in \S \ref{subsec:auto-power}, therefore isolating the power due to galaxy clustering. This term can be described by Limber's equation \citep{Limber1953}

\begin{equation}
P^{\textrm{Cl}}_{IR}(q) = \int_{z_{min}}^{z_{max}} \frac{H(z)}{cd_c^2(z)}\left[\frac{dF_{CIB}}{dz}\right]^2 P(qd_c^{-1},z)dz
\label{eq:limber_new}
\end{equation}
where $H$ is the Hubble factor, $d_c$ is the co-moving distance, $dF_{CIB}/dz$ is the flux production rate of the CIB, and $P(qd_c^{-1},z)$ is the 3D power spectrum. The flux production rate of the CIB can be written as
\begin{equation}
\frac{dF_{CIB}}{dz} = \int f(m_{AB}) \frac{d^3N}{dm_{AB}d\Omega dz} dm_{AB}.
\label{eq:flux_prod_rate_new}
\end{equation}
Furthermore, for our analysis we only consider angular scales $\gtrsim 300''$ as in this regime the galaxy power spectrum is safely within the linear approximation \citep{helgason12}. As a result the 3D galaxy power spectrum is defined as
\begin{equation}
P(qd_c^{-1},z)=b^2 P_{\Lambda \textrm{CDM}}(k,z)
\label{eq:3D_ps}
\end{equation}
where $b$ is the large-scale bias, or clustering strength of galaxies with respect to dark matter, and $ P_{\Lambda \textrm{CDM}}$ is the 3D power spectrum of dark matter. Essentially, the bias parameter encapsulates the physics of galaxy formation and evolution, including processes impacting and/or driven by star formation and black hole activity \citep{Desjacques2018,Paquereau2025}. To compute the linear matter power spectrum, we use the cosmology package \texttt{COLOSSUS} \citep{Diemer2018}.


In this analysis the sole parameter we fit is the average biasing of the CIB sources, $\tilde{b}_{IR}$. In this fit we employ the \texttt{Python} package \texttt{emcee}, an affine-invariant ensemble sampler introduced by \cite{ForemanMackey2013}. We sample parameter-space assuming a flat prior distribution using 5 walkers and 500 steps. The likelihood function $\textrm{ln}(P)$ is written in the form of 
\begin{equation}
\textrm{ln}(P)=-\frac{1}{2}\chi^2
\end{equation}
where
\begin{equation}
\chi^2=\sum \left(\frac{P^{\textrm{tot}}_{IR}(q)-P^{\textrm{obs}}_{IR}(q)}{\sigma^{\textrm{obs}}_{IR}(q)}\right)^2
\end{equation}
where the ``tot" superscript indicates the model power spectra while the ``obs" superscript indicates the observed power spectra. We confirm that in each redshift interval the MCMC chain converges and is independent of the prior bounds used. {The  best-fit values of $\tilde{b}_{IR}$ in the [0 -- 3], [3 -- 6], and [6 -- 13] $\Delta z$ bins are $1.03_{-0.10}^{+0.10}$, $3.21_{-0.25}^{+0.24}$, and $6.51_{-0.84}^{+0.77}$ respectively. Each fit has a reduced-$\chi^2$ of 1.33, 0.80, and 0.83 for each redshift interval respectively. The fitted CIB auto-power spectra and associated $\tilde{b}_{IR}$ values can be found in Fig. \ref{fig:aps_bias_fits}.}

\subsection{Estimated CXB Flux Production}\label{subsec:cxb}
An in-depth treatment of the CXB in regards to its population characteristics, abundance, and clustering will be the focus of a forthcoming paper. However, we can first estimate the CXB flux production rate as a function of redshift using the auto- and cross-power spectra computed in this study. Here, we first approximate the \textit{coherent} rms CXB fluctuations $\delta F_{CXB}$ as a function of $q$ in each $\Delta z$ bin to be
\begin{equation}
\delta F_{CXB}(q) = \sqrt{\frac{q^2}{2\pi}\frac{P_{IR,X}^2(q)}{P_{IR}(q)P_{\textrm{PSF}}(q)}}
\end{equation}
where we introduce $P_{\textrm{PSF}}(q)$ to be the power due to the \textit{Chandra} beam, taken to be Eq. 11 of \cite{Cappelluti2012}. We use Monte Carlo sampling with the associated errors of the auto- and cross-power spectra (with a total of 10000 samples) to compute the mean flux of the CXB. This procedure is done in each $\Delta z$ bin. The resulting CXB flux $\langle F_{CXB} \rangle$ in each interval is $9.68_{-4.15}^{+5.73}$, $8.72_{-3.83}^{+5.54}$, and $7.99_{-3.52}^{+5.24}$ $\times 10^{-14}$ erg/s/cm$^2$/deg$^2$ respectively. We find the total \textit{coherent} CXB flux $\langle F_{CXB} \rangle$ to be $2.64_{-0.67}^{+0.52}$ $\times 10^{-13}$ erg/s/cm$^2$/deg$^2$, which is within the upper limit of the total unresolved CXB flux measured in \cite{Cappelluti2017cxb}.

\begin{figure}[ht!]
\begin{center}
    \includegraphics[width = \columnwidth]{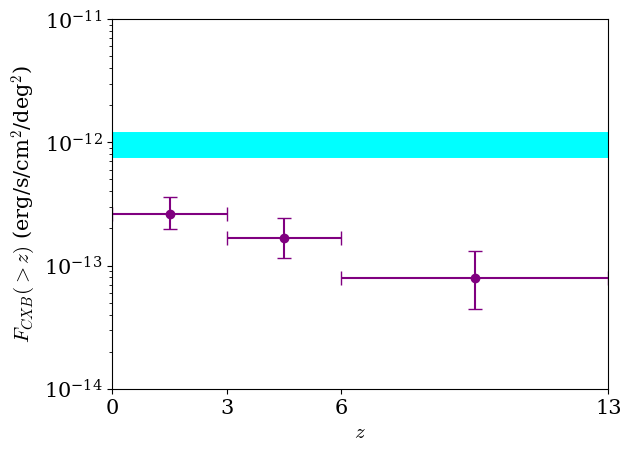}
\end{center}
\caption{Estimated Cumulative flux contribution of the CXB as a function of redshift. The cyan bar at $0.97 \times 10^{-12}$ erg/s/cm$^{2}$/deg$^{2}$ is the upper limit placed on the unresolved CXB in \cite{Cappelluti2017cxb}. The flux contribution of the CXB gets progressively fainter as a function of redshift as expected \citep{Helgason2014,Cappelluti2017cxb}, albeit with a weak decline. Errors are reported at the $1\sigma$ confidence level.}
\label{fig:flux_prod_rate}
\end{figure}

In Figure~\ref{fig:flux_prod_rate} we show the cumulative [0.5-2] keV CXB flux production as a function of redshift, compared with the flux derived by \cite{Cappelluti2017cxb} in the COSMOS-Legacy Survey for sources below IR magnitude limits similar to those applied in our mask. \cite{Cappelluti2017cxb} reported an unresolved CXB surface brightness of \(9.7^{+1.6}_{-1.8} \times 10^{-13}\) erg/s/cm\(^2\)/deg\(^2\), whereas we find a value of \(2.63_{-0.67}^{+0.52}\) erg/s/cm\(^2\)/deg\(^2\), corresponding to \(27^{+13}_{-10}\%\) of their unresolved CXB. These results show that approximately 94\% of the CXB has been resolved. When accounting for systematic uncertainties in the total CXB measurements, our findings are consistent within 2$\sigma$ of fully resolving the extragalactic [0.5-2] keV CXB.

Without fitting the CIB-CXB cross-power, we can still estimate lower limits on the biasing of X-ray sources, $\tilde{b}_X$. This is done by using the best-fit $\tilde{b}_{IR}$, in addition to the derived flux production rates of the CIB and CXB. Then using the following equation
{
\begin{align}
P^{\textrm{Cl}}_{IR,X}(q) &= \int_{z_{min}}^{z_{max}} \frac{H(z)}{c\,d_c^2(z)} \frac{dF_{CIB}}{dz}\frac{dF_{CXB}}{dz} \nonumber \\
&\quad \times \tilde{b}_{IR}\tilde{b}_{X}P_{\Lambda \textrm{CDM}}(q\,d_c^{-1},z)dz
\label{eq:limber_new_cross}
\end{align}}
{a rough estimate for $\tilde{b}_X$ at $2\pi/q \sim 1000''$ is determined. Because of the relatively lower signal in the $\Delta  z = [3-6]$ bin at larger angular scales, we only perform this estimate in the low-$z$ and high-$z$ bins. For the former, we find a value of $\sim 1.93$ while in the latter an estimate of $\sim 7.71$ (see Fig. \ref{fig:aps_bias_fits}).}

\subsection{Low-$z$ Populations}
The best-fit $\tilde{b}_{IR}$ of {$1.03_{-0.10}^{+0.10}$ in the [0 -- 3] redshift bin is consistent with previously computed values of the large-scale biasing of SFGs \citep{Coil2017,Hale2018}. Using the bias prescription from \cite{SMT2001}, this best-fit value corresponds to a characteristic halo mass $M_h$ of $\sim 10^{10}$ $M_{\odot} h^{-1}$}, further pointing toward a population of faint SFGs resolved by $JWST$. This is expected given that the majority of the sources in the COSMOS-Web field are SFGs \citep{Shuntov2024}. Additionally, the $\tilde{b}_{IR}$ value discussed here is consistent with the galaxy bias for SFGs computed in the COSMOS-Web field \citep{Paquereau2025}, with their results ranging from $b\sim 1-5$ in the redshift interval [0 -- 3], with the spread due to binning galaxies according to their stellar mass $M_{\star}$. The total z-PDF of our sources peaks near $z\sim1$, where the bias reported in \cite{Paquereau2025} is $\sim 1.5$, bringing our results into even closer agreement. Furthermore, these results are unsurprising given the apparent consistency of these results with the CIB auto-power spectrum modeled by \cite{helgason12}, which is entirely the result of stellar emission produced mostly within the range $1 \lesssim z \lesssim 3$.

The low-$z$ CIB signal detected in this study has several implications for the $\sim 5\sigma$ cross-power signal reported. It is expected that SFGs (1) dominate the low-flux regime of the X-ray log($N$)-log($S$) and (2) contribute to $\sim 25\%$ of the unresolved CXB anisotropies \citep{Cappelluti2012,Cappelluti2016}. As a result, it is entirely possible that the coherence seen at low-$z$ can at least in part be due to SFGs that emit in both the IR and soft X-rays \citep{Helgason2014}.

Additionally, we expect that star-forming populations and AGNs cluster within the same large-scale structures \citep{Helgason2014}, so the CIB-CXB cross-power signal detected may be a reflection of these distinct populations clustered together. The lower limit of $\tilde{b}_X$ at low-$z$ is significantly higher than the biasing of CIB sources, which may support the claim that the cross-power signal is produced by SFGs in the IR and AGNs in the X-ray. { As $\tilde{b}_X \sim 1.93$, this corresponds to $M_h \sim 10^{12}$ $M_{\odot}h^{-1}$, a halo mass regime characteristic of AGN hosts. According to \cite{Helgason2014}, between $0 \lesssim z \lesssim 3$ the biasing of AGNs are as low as $b\sim 1.5$ and can go as high as $b\sim 4$, which is consistent with our value.} Further modeling of the CIB-CXB cross-power signal is needed to shed light on the precise populations contributing to the unresolved soft CXB in this redshift range.

\subsection{High-$z$ Populations} {The best-fit high-$z$ $\tilde{b}_{IR}$ and $\langle F_{CXB} \rangle$ values of $6.51_{-0.84}^{+0.77}$ and $7.99_{-3.52}^{+5.24}$ $\times 10^{-14}$ erg s$^{-1}$cm$^{-2}$deg$^{-2}$ respectively hold several implications for the sources contributing to the CIB and CIB-CXB angular power spectra. The $\tilde{b}_{IR}$ derived in this redshift interval is within the $\sim 7-15$ range put forth by \cite{Paquereau2025}, and corresponds to a halo mass of $\sim 10^{10}$ $M_{\odot}h^{-1}$.} According to the results of \cite{Shuntov2024}, this halo mass corresponds to galaxies with stellar masses as high as $\sim 5\times10^{10}$ $M_{\odot}$. It is entirely possible that these early, massive galaxies were clustered in the same large-scale environments as early AGNs. {The lower limit we place on the $\tilde{b}_X$ parameter at high-$z$ of $\sim 7.71$ (corresponding to $M_{h}\sim10^{10}$ $M_{\odot}h^{-1}$)} is consistent with previously modeled high-mass SMBH seeds \citep{Ricarte2019}, suggesting that these sources may be massively accreting SMBHs. Further on this note, it has been proposed that these high-$z$ AGNs can boost the CIB-CXB cross-power signal \citep{Ricarte2019,Cappelluti2022}, an interpretation that can be tested with the $\sim 5\sigma$ signal reported in this work. If we assume that the CXB flux at $z>6$ is primarily powered by accretion onto SMBHs \citep{Cappelluti2022}, we can invoke Sołtan's Argument \citep{Soltan1982} to estimate the BH accreted mass density $\rho_{acc}$ as outlined in \cite{Salvaterra2012}
\begin{equation}
\begin{aligned}
\rho_{\text{acc}}(z) = 4\pi \frac{1-\epsilon}{\epsilon c^3} \frac{E_0 J_{E_0}}{f_X (1-\alpha)} \frac{\alpha + \gamma + \frac{3}{2}}{\gamma + \frac{3}{2}} (1+z)^\alpha \\
\quad \times \left[\left(\frac{E_M}{E_0}\right)^{1-\alpha} - \left(\frac{E_m}{E_0}\right)^{1-\alpha}\right]
\end{aligned}
\end{equation}
where $\epsilon$, $f_X$, $\alpha$, $\gamma$ are the mass-accretion conversion efficiency, bolometric correction constant, AGN spectrum slope, and a constant characterizing the redshift evolution of the comoving specific emissivity respectively. $E_0$ is the background energy and $J_{E_0}$ is the emissivity observed at $E_0$, and $E_m$ and $E_M$ are the lower and upper bounds (respectively) of the X-ray band evaluated. We use the bolometric correction constant at z = 6 derived in \cite{Ricarte2019} and assume that $(\alpha + \gamma + \frac{3}{2})/(\gamma + \frac{3}{2}) \simeq 1$. Additionally, we assume $\epsilon$ to be 0.1. Plugging in each of these values and the derived CXB flux at $z > 6$, we estimate that $\rho_{acc} \approx 10^{5.15}$ $M_{\odot}$ Mpc$^{-3}$. Our result is consistent with the upper limits put forth in \cite{Cappelluti2017cxb}. It is also important to include that our $\rho_{acc}$ at $z=6$ is only slightly under local estimates according to  \cite{Hopkins2007} and \cite{Shankar2009} which are log$\rho_{acc}(z=0)\sim5.5-5.7 $. This is due in-part by the assumed bolometric correction from \cite{Ricarte2019} and the order-of-magnitude CXB flux estimate here.

Another interesting estimate is the number density of sources producing the level of Cosmic X-ray Background (CXB) responsible for the observed cross-power. By assuming, for simplicity, that the X-ray source counts in the high-redshift bin (i.e., $\Delta z =$ [6 -- 13]) behave in a Euclidean manner, the CXB production at redshift $z$ can be expressed as
\begin{equation}
F_{\rm CXB}(z) = \int_{S_{lim}}^{\infty} S\,\frac{dN}{dS}(z)\,dS
\end{equation}
where the differential number counts can be written as
\begin{equation}
\frac{dN}{dS}(z) \propto k\left(\frac{S}{S_0}\right)^{-2.5}.
\end{equation}
Here, $k$ is the normalization and \(S_0 = 4.2 \times 10^{-18}\) erg s\(^{-1}\) cm\(^{-2}\) \citep[the typical flux limit of a moderately deep AXIS survey,][]{axis}. With this assumption, we find a source density on the order of \(500\)–\(700\) deg\(^{-2}\), consistent with the predictions of the semi-analytic model of \citet{ricarte18}. Although the number density can be constrained, we cannot disentangle their SMBH seeding models.

{ The CXB flux and biasing estimates derived in this work can be placed in the broader context of the recently discovered ``Little Red Dots" (or LRDs) by JWST. These objects are found at $z>4$, and are AGNs with compact, red hosts \citep{Harikane2023,Kocevski2023}. Since their discovery, the nature of LRDs have been hotly debated with the two prevailing explanations, being that (1) they are highly obscured and/or (2) undergo super-Eddington accretion \cite{Lambrides2024,Pacucci2024}. X-ray emitting LRDs that have not been resolved by \textit{Chandra}, if abundant enough, may have a measurable contribution to the CIB-CXB power spectrum. A signal such as this would have important applications for studying the physics of LRD hosts and their evolution as a function of redshift.}

Future work will leverage the robust data obtained in the COSMOS-Web field, including a catalog of star formation rates (SFRs) and stellar masses ($M_{\star}$) (priv. comm.) computed using the SED fitting code \texttt{CIGALE} \citep{Boquien2019}. These quantities alongside others will be used in conjunction with the artificial map creation method invoked in this work to shed light on the contributions of SFGs and AGNs to the observed CIB-CXB cross-power signal at high-$z$.




\section{Summary}
In this study, we utilize \textit{JWST} observations of resolved sources in the F277W and F444W bands from the COSMOS-Web survey \citep{Casey2023} to create CIB fluctuation maps. Sources brighter than the magnitude limit $m_{AB} = 25$ (the \textit{Spitzer} limit) were masked. The auto-power spectra of each map were computed and cross-correlated with the unresolved [0.5-2] keV CXB as observed by \textit{Chandra}. Our main results can be summarized as the following: 
\begin{enumerate}
  \item The CIB-CXB cross-power signal measured with sources below the \textit{Spitzer} flux limit is statistically significant with a $S/N$ of 4.80 and 6.20 at angular scales $1-1000''$, for the F277W and F444W CIB fluctuations respectively. The cross-power spectrum at F444W is consistent with the measured spectra from \cite{Cappelluti2013} and \cite{Li2018}, demonstrating that the previously detected cross-power can be explained by galaxy populations revealed by \textit{JWST}. This significance allows us to conclude that sources resolved by \textit{JWST} emit soft X-rays and are clustered on larger spatial scales. The significance of this cross-power spectrum in different $\Delta z$ ranges has also been computed and notably we find for the first time evidence of significant emission of X-rays among the newly discovered population of \textit{JWST} $z > 6$ galaxies.  
  \item We estimate the lower limit of the unresolved [0.5-2] keV CXB flux in the redshift range $0<z<13$ to be $2.64_{-0.67}^{+0.52}$ $\times 10^{-13}$ erg/s/cm$^2$/deg$^2$. If the CXB flux at $z > 6$ is produced entirely by AGNs, this corresponds to a black hole accreted mass density of $\rho_{acc} \approx 10^{5.15}$ $M_{\odot}$ Mpc$^{-3}$. These results are consistent with the findings of \citep{Cappelluti2017cxb}. {In the low- and high-$z$ intervals probed we place lower limits on the biasing of these X-ray sources to be $\sim$ 1.93 and 7.71 respectively, corresponding to halo masses of $10^{12}$ and $10^{10}$ $M_{\odot}h^{-1}$}.
  \item{We resolve \(27^{+13}_{-10}\%\) of the unresolved CXB measured by \cite{Cappelluti2017cxb}. This brings the resolved fraction of the CXB to at least $\sim 94\%$. The remaining fraction of the unresolved CXB, if any, can be attributed to diffuse X-ray sources and/or systematic uncertainties in the actual CXB flux. At \(6 < z < 13\) the CXB production rate measured here roughly corresponds to a source surface density of \(500\)–\(700\) deg\(^{-2}\).}
  \item We find a statistically significant excess in the CIB auto-power spectrum of maps in both the F277W and F444W maps. For the latter, we find that the auto-power spectrum is significantly lower than the 4.5 $\mu m$ CIB auto-power spectrum computed in \cite{Kashlinsky2012} and \cite{Li2018}, leaving their detected clustering component still unexplained by known galaxy populations \citep{helgason12}. 
  \item The CIB power spectrum was evaluated at $\Delta z$= [0 -- 3], [3 -- 6], [6 -- 13], and [0 -- 13]. We find a clustering excess at large scales with fluctuations that drop as redshift increases. Specifically, with the auto-power spectra in the [0 -- 3] and [3 -- 6] bins, we find close similarity with the reconstructed CIB fluctuations derived in \cite{helgason12}. {Each auto-power spectrum is fitted using MCMC methods to obtain large-scale biases of $1.03_{-0.10}^{+0.10}$, $3.21_{-0.25}^{+0.24}$, and $6.51_{-0.84}^{+0.77}$ for each respective $\Delta z$ bin. These values correspond to halo masses on the order of $\sim 10^{10}$ $M_{\odot}h^{-1}$.}\\
\end{enumerate}

The unprecedented quality data gathered in the COSMOS-Web field provides a powerful boost to the Fourier analysis of the very faint CIB fluctuations as a function of source brightness and redshift. Future work will make use of derived source properties in order to further constrain the populations behind the CIB-CXB cross-power spectrum at all spatial scales, quantifying the abundance and clustering of SFGs and AGNs. This will have important implications for constraining SMBH seeding mechanisms and populations responsible for reionization. Finally, this study shows that, although individually undetectable, high-$z$ AGN contribute to the CXB. Consequently, future, more powerful X-ray telescopes, such as AXIS \citep{axis}, will resolve these sources and further enhance our understanding of faint, high-$z$ accreting SMBHs.

\begin{acknowledgments}
{ We thank the anonymous referee for their insightful comments which greatly improved the paper.} {A.J.K., N.C.}, and J.S. acknowledge JWST-AR-03303.002-A and Chandra grant AR3-24008X.{ A.J.K., N.C.}, and J.S. acknowledge the University of Miami for partially supporting this investigation.{ A.J.K.} acknowledges insightful discussions with Alberto Magaraggia, Angelo Ricarte, and Richard G. Arendt regarding map-making and X-ray number counts. {A.J.K. also thanks Louise Paquereau for assisting in improving the large-scale bias models.} 

The scientific results reported in this article are based on observations made by the Chandra X-ray Observatory and published previously in \cite{Civano2016}.
All the authors acknowledge the outstanding job of the Chandra X-ray Observatory and JWST teams. We kindly acknowledge the COSMOS-Web team for insightful discussions and support on data products.
\end{acknowledgments}

%

\vspace{5mm}
\facilities{JWST, Chandra}


\software{astropy \citep{2013A&A...558A..33A,2018AJ....156..123A},  
          Source Extractor \citep{1996A&AS..117..393B}
          COLOSSUS \citep{Diemer2018}, emcee \citep{ForemanMackey2013}}

\bibliography{sample631}{}

\begin{thebibliography}{}
\expandafter\ifx\csname natexlab\endcsname\relax\def\natexlab#1{#1}\fi
\providecommand{\url}[1]{\href{#1}{#1}}
\providecommand{\dodoi}[1]{doi:~\href{http://doi.org/#1}{\nolinkurl{#1}}}
\providecommand{\doeprint}[1]{\href{http://ascl.net/#1}{\nolinkurl{http://ascl.net/#1}}}
\providecommand{\doarXiv}[1]{\href{https://arxiv.org/abs/#1}{\nolinkurl{https://arxiv.org/abs/#1}}}

\bibitem[{{Allevato} {et~al.}(2016){Allevato}, {Civano}, {Finoguenov}, {Marchesi}, {Shankar}, {Zamorani}, {Hasinger}, {Salvato}, {Miyaji}, {Gilli}, {Cappelluti}, {Brusa}, {Suh}, {Lanzuisi}, {Trakhtenbrot}, {Griffiths}, {Vignali}, {Schawinski}, \& {Karim}}]{Allevato2016}
{Allevato}, V., {Civano}, F., {Finoguenov}, A., {et~al.} 2016, \apj, 832, 70, \dodoi{10.3847/0004-637X/832/1/70}

\bibitem[{{Arnouts} {et~al.}(1999){Arnouts}, {Cristiani}, {Moscardini}, {Matarrese}, {Lucchin}, {Fontana}, \& {Giallongo}}]{Arnouts1999}
{Arnouts}, S., {Cristiani}, S., {Moscardini}, L., {et~al.} 1999, \mnras, 310, 540, \dodoi{10.1046/j.1365-8711.1999.02978.x}

\bibitem[{{Astropy Collaboration} {et~al.}(2013){Astropy Collaboration}, {Robitaille}, {Tollerud}, {Greenfield}, {Droettboom}, {Bray}, {Aldcroft}, {Davis}, {Ginsburg}, {Price-Whelan}, {Kerzendorf}, {Conley}, {Crighton}, {Barbary}, {Muna}, {Ferguson}, {Grollier}, {Parikh}, {Nair}, {Unther}, {Deil}, {Woillez}, {Conseil}, {Kramer}, {Turner}, {Singer}, {Fox}, {Weaver}, {Zabalza}, {Edwards}, {Azalee Bostroem}, {Burke}, {Casey}, {Crawford}, {Dencheva}, {Ely}, {Jenness}, {Labrie}, {Lim}, {Pierfederici}, {Pontzen}, {Ptak}, {Refsdal}, {Servillat}, \& {Streicher}}]{2013A&A...558A..33A}
{Astropy Collaboration}, {Robitaille}, T.~P., {Tollerud}, E.~J., {et~al.} 2013, \aap, 558, A33, \dodoi{10.1051/0004-6361/201322068}

\bibitem[{{Astropy Collaboration} {et~al.}(2018){Astropy Collaboration}, {Price-Whelan}, {Sip{\H{o}}cz}, {G{\"u}nther}, {Lim}, {Crawford}, {Conseil}, {Shupe}, {Craig}, {Dencheva}, {Ginsburg}, {VanderPlas}, {Bradley}, {P{\'e}rez-Su{\'a}rez}, {de Val-Borro}, {Aldcroft}, {Cruz}, {Robitaille}, {Tollerud}, {Ardelean}, {Babej}, {Bach}, {Bachetti}, {Bakanov}, {Bamford}, {Barentsen}, {Barmby}, {Baumbach}, {Berry}, {Biscani}, {Boquien}, {Bostroem}, {Bouma}, {Brammer}, {Bray}, {Breytenbach}, {Buddelmeijer}, {Burke}, {Calderone}, {Cano Rodr{\'\i}guez}, {Cara}, {Cardoso}, {Cheedella}, {Copin}, {Corrales}, {Crichton}, {D'Avella}, {Deil}, {Depagne}, {Dietrich}, {Donath}, {Droettboom}, {Earl}, {Erben}, {Fabbro}, {Ferreira}, {Finethy}, {Fox}, {Garrison}, {Gibbons}, {Goldstein}, {Gommers}, {Greco}, {Greenfield}, {Groener}, {Grollier}, {Hagen}, {Hirst}, {Homeier}, {Horton}, {Hosseinzadeh}, {Hu}, {Hunkeler}, {Ivezi{\'c}}, {Jain}, {Jenness}, {Kanarek}, {Kendrew}, {Kern}, {Kerzendorf}, {Khvalko}, {King}, {Kirkby}, {Kulkarni},
  {Kumar}, {Lee}, {Lenz}, {Littlefair}, {Ma}, {Macleod}, {Mastropietro}, {McCully}, {Montagnac}, {Morris}, {Mueller}, {Mumford}, {Muna}, {Murphy}, {Nelson}, {Nguyen}, {Ninan}, {N{\"o}the}, {Ogaz}, {Oh}, {Parejko}, {Parley}, {Pascual}, {Patil}, {Patil}, {Plunkett}, {Prochaska}, {Rastogi}, {Reddy Janga}, {Sabater}, {Sakurikar}, {Seifert}, {Sherbert}, {Sherwood-Taylor}, {Shih}, {Sick}, {Silbiger}, {Singanamalla}, {Singer}, {Sladen}, {Sooley}, {Sornarajah}, {Streicher}, {Teuben}, {Thomas}, {Tremblay}, {Turner}, {Terr{\'o}n}, {van Kerkwijk}, {de la Vega}, {Watkins}, {Weaver}, {Whitmore}, {Woillez}, {Zabalza}, \& {Astropy Contributors}}]{2018AJ....156..123A}
{Astropy Collaboration}, {Price-Whelan}, A.~M., {Sip{\H{o}}cz}, B.~M., {et~al.} 2018, \aj, 156, 123, \dodoi{10.3847/1538-3881/aabc4f}

\bibitem[{{Bertin} \& {Arnouts}(1996)}]{1996A&AS..117..393B}
{Bertin}, E., \& {Arnouts}, S. 1996, \aaps, 117, 393, \dodoi{10.1051/aas:1996164}

\bibitem[{{Bertin} {et~al.}(2020){Bertin}, {Schefer}, {Apostolakos}, {{\'A}lvarez-Ayll{\'o}n}, {Dubath}, \& {K{\"u}mmel}}]{Bertin2020}
{Bertin}, E., {Schefer}, M., {Apostolakos}, N., {et~al.} 2020, in Astronomical Society of the Pacific Conference Series, Vol. 527, Astronomical Data Analysis Software and Systems XXIX, ed. R.~{Pizzo}, E.~R. {Deul}, J.~D. {Mol}, J.~{de Plaa}, \& H.~{Verkouter}, 461

\bibitem[{{Bogd{\'a}n} {et~al.}(2024){Bogd{\'a}n}, {Goulding}, {Natarajan}, {Kov{\'a}cs}, {Tremblay}, {Chadayammuri}, {Volonteri}, {Kraft}, {Forman}, {Jones}, {Churazov}, \& {Zhuravleva}}]{2024NatAs...8..126B}
{Bogd{\'a}n}, {\'A}., {Goulding}, A.~D., {Natarajan}, P., {et~al.} 2024, Nature Astronomy, 8, 126, \dodoi{10.1038/s41550-023-02111-9}

\bibitem[{{Boquien} {et~al.}(2019){Boquien}, {Burgarella}, {Roehlly}, {Buat}, {Ciesla}, {Corre}, {Inoue}, \& {Salas}}]{Boquien2019}
{Boquien}, M., {Burgarella}, D., {Roehlly}, Y., {et~al.} 2019, \aap, 622, A103, \dodoi{10.1051/0004-6361/201834156}

\bibitem[{{Capak} {et~al.}(2016){Capak}, {Arendt}, {Arnouts}, {Bartlett}, {Bouwens}, {Brinchman}, {Brodwin}, {Carollo}, {Castander}, {Charlot}, {Chary}, {Cohen}, {Cooray}, {Conselice}, {Coupon}, {Cuby}, {Culliandre}, {Davidzon}, {Dole}, {Dunlop}, {Eisenhardt}, {Ferrara}, {Gardner}, {Hasinger}, {Hildebrandt}, {Ho}, {Ilbert}, {Jouvel}, {Kashlinsky}, {LeFevre}, {LeFloc'h}, {Maraston}, {Masters}, {McCracken}, {Mei}, {Mellier}, {Mitchell-Wynn}, {Moustakas}, {Nayyeri}, {Paltani}, {Rhodes}, {Salvato}, {Sanders}, {Scaramella}, {Scarlata}, {Scoville}, {Silverman}, {Speagle}, {Stanford}, {Stern}, {Teplitz}, \& {Toft}}]{Capak2016}
{Capak}, P., {Arendt}, R., {Arnouts}, S., {et~al.} 2016, {The Euclid/WFIRST Spitzer Legacy Survey}, Spitzer Proposal ID \#13058

\bibitem[{{Cappelluti} {et~al.}(2022){Cappelluti}, {Hasinger}, \& {Natarajan}}]{Cappelluti2022}
{Cappelluti}, N., {Hasinger}, G., \& {Natarajan}, P. 2022, \apj, 926, 205, \dodoi{10.3847/1538-4357/ac332d}

\bibitem[{{Cappelluti} {et~al.}(2012{\natexlab{a}}){Cappelluti}, {Ranalli}, {Roncarelli}, {Arevalo}, {Zamorani}, {Comastri}, {Gilli}, {Rovilos}, {Vignali}, {Allevato}, {Finoguenov}, {Miyaji}, {Nicastro}, {Georgantopoulos}, \& {Kashlinsky}}]{cap12}
{Cappelluti}, N., {Ranalli}, P., {Roncarelli}, M., {et~al.} 2012{\natexlab{a}}, \mnras, 427, 651, \dodoi{10.1111/j.1365-2966.2012.21867.x}

\bibitem[{{Cappelluti} {et~al.}(2012{\natexlab{b}}){Cappelluti}, {Ranalli}, {Roncarelli}, {Arevalo}, {Zamorani}, {Comastri}, {Gilli}, {Rovilos}, {Vignali}, {Allevato}, {Finoguenov}, {Miyaji}, {Nicastro}, {Georgantopoulos}, \& {Kashlinsky}}]{Cappelluti2012}
---. 2012{\natexlab{b}}, \mnras, 427, 651, \dodoi{10.1111/j.1365-2966.2012.21867.x}

\bibitem[{{Cappelluti} {et~al.}(2013){Cappelluti}, {Kashlinsky}, {Arendt}, {Comastri}, {Fazio}, {Finoguenov}, {Hasinger}, {Mather}, {Miyaji}, \& {Moseley}}]{Cappelluti2013}
{Cappelluti}, N., {Kashlinsky}, A., {Arendt}, R.~G., {et~al.} 2013, \apj, 769, 68, \dodoi{10.1088/0004-637X/769/1/68}

\bibitem[{{Cappelluti} {et~al.}(2016){Cappelluti}, {Comastri}, {Fontana}, {Zamorani}, {Amorin}, {Castellano}, {Merlin}, {Santini}, {Elbaz}, {Schreiber}, {Shu}, {Wang}, {Dunlop}, {Bourne}, {Bruce}, {Buitrago}, {Micha{\l}owski}, {Derriere}, {Ferguson}, {Faber}, \& {Vito}}]{Cappelluti2016}
{Cappelluti}, N., {Comastri}, A., {Fontana}, A., {et~al.} 2016, \apj, 823, 95, \dodoi{10.3847/0004-637X/823/2/95}

\bibitem[{{Cappelluti} {et~al.}(2017{\natexlab{a}}){Cappelluti}, {Li}, {Ricarte}, {Agarwal}, {Allevato}, {Tasnim Ananna}, {Ajello}, {Civano}, {Comastri}, {Elvis}, {Finoguenov}, {Gilli}, {Hasinger}, {Marchesi}, {Natarajan}, {Pacucci}, {Treister}, \& {Urry}}]{Cappelluti2017cxb}
{Cappelluti}, N., {Li}, Y., {Ricarte}, A., {et~al.} 2017{\natexlab{a}}, \apj, 837, 19, \dodoi{10.3847/1538-4357/aa5ea4}

\bibitem[{{Cappelluti} {et~al.}(2017{\natexlab{b}}){Cappelluti}, {Arendt}, {Kashlinsky}, {Li}, {Hasinger}, {Helgason}, {Urry}, {Natarajan}, \& {Finoguenov}}]{Cappelluti2017}
{Cappelluti}, N., {Arendt}, R., {Kashlinsky}, A., {et~al.} 2017{\natexlab{b}}, \apjl, 847, L11, \dodoi{10.3847/2041-8213/aa8acd}

\bibitem[{{Casey} {et~al.}(2023){Casey}, {Kartaltepe}, {Drakos}, {Franco}, {Harish}, {Paquereau}, {Ilbert}, {Rose}, {Cox}, {Nightingale}, {Robertson}, {Silverman}, {Koekemoer}, {Massey}, {McCracken}, {Rhodes}, {Akins}, {Allen}, {Amvrosiadis}, {Arango-Toro}, {Bagley}, {Bongiorno}, {Capak}, {Champagne}, {Chartab}, {Ch{\'a}vez Ortiz}, {Chworowsky}, {Cooke}, {Cooper}, {Darvish}, {Ding}, {Faisst}, {Finkelstein}, {Fujimoto}, {Gentile}, {Gillman}, {Gould}, {Gozaliasl}, {Hayward}, {He}, {Hemmati}, {Hirschmann}, {Jahnke}, {Jin}, {Khostovan}, {Kokorev}, {Lambrides}, {Laigle}, {Larson}, {Leung}, {Liu}, {Liaudat}, {Long}, {Magdis}, {Mahler}, {Mainieri}, {Manning}, {Maraston}, {Martin}, {McCleary}, {McKinney}, {McPartland}, {Mobasher}, {Pattnaik}, {Renzini}, {Rich}, {Sanders}, {Sattari}, {Scognamiglio}, {Scoville}, {Sheth}, {Shuntov}, {Sparre}, {Suzuki}, {Talia}, {Toft}, {Trakhtenbrot}, {Urry}, {Valentino}, {Vanderhoof}, {Vardoulaki}, {Weaver}, {Whitaker}, {Wilkins}, {Yang}, \& {Zavala}}]{Casey2023}
{Casey}, C.~M., {Kartaltepe}, J.~S., {Drakos}, N.~E., {et~al.} 2023, \apj, 954, 31, \dodoi{10.3847/1538-4357/acc2bc}

\bibitem[{{Civano} {et~al.}(2016){Civano}, {Marchesi}, {Comastri}, {Urry}, {Elvis}, {Cappelluti}, {Puccetti}, {Brusa}, {Zamorani}, {Hasinger}, {Aldcroft}, {Alexander}, {Allevato}, {Brunner}, {Capak}, {Finoguenov}, {Fiore}, {Fruscione}, {Gilli}, {Glotfelty}, {Griffiths}, {Hao}, {Harrison}, {Jahnke}, {Kartaltepe}, {Karim}, {LaMassa}, {Lanzuisi}, {Miyaji}, {Ranalli}, {Salvato}, {Sargent}, {Scoville}, {Schawinski}, {Schinnerer}, {Silverman}, {Smolcic}, {Stern}, {Toft}, {Trakhtenbrot}, {Treister}, \& {Vignali}}]{Civano2016}
{Civano}, F., {Marchesi}, S., {Comastri}, A., {et~al.} 2016, \apj, 819, 62, \dodoi{10.3847/0004-637X/819/1/62}

\bibitem[{{Coil} {et~al.}(2017){Coil}, {Mendez}, {Eisenstein}, \& {Moustakas}}]{Coil2017}
{Coil}, A.~L., {Mendez}, A.~J., {Eisenstein}, D.~J., \& {Moustakas}, J. 2017, \apj, 838, 87, \dodoi{10.3847/1538-4357/aa63ec}

\bibitem[{{Cooray} {et~al.}(2012){Cooray}, {Smidt}, {de Bernardis}, {Gong}, {Stern}, {Ashby}, {Eisenhardt}, {Frazer}, {Gonzalez}, {Kochanek}, {Koz{\l}owski}, \& {Wright}}]{Cooray2012}
{Cooray}, A., {Smidt}, J., {de Bernardis}, F., {et~al.} 2012, \nat, 490, 514, \dodoi{10.1038/nature11474}

\bibitem[{{Desjacques} {et~al.}(2018){Desjacques}, {Jeong}, \& {Schmidt}}]{Desjacques2018}
{Desjacques}, V., {Jeong}, D., \& {Schmidt}, F. 2018, \physrep, 733, 1, \dodoi{10.1016/j.physrep.2017.12.002}

\bibitem[{{Diemer}(2018)}]{Diemer2018}
{Diemer}, B. 2018, \apjs, 239, 35, \dodoi{10.3847/1538-4365/aaee8c}

\bibitem[{{Dunlop} {et~al.}(2021){Dunlop}, {Abraham}, {Ashby}, {Bagley}, {Best}, {Bongiorno}, {Bouwens}, {Bowler}, {Brammer}, {Bremer}, {Calabro'}, {Carnall}, {Castellano}, {Cirasuolo}, {Conselice}, {Cullen}, {Dave}, {Dayal}, {Dekel}, {Dickinson}, {Duncan}, {Elbaz}, {Ellis}, {Ferguson}, {Ferrara}, {Finkelstein}, {Fontana}, {Furlanetto}, {Fynbo}, {Gallerani}, {Gardner}, {Giavalisco}, {Grazian}, {Grogin}, {Harikane}, {Hopkins}, {Ilbert}, {Illingworth}, {Juneau}, {Jung}, {Kartaltepe}, {Kassin}, {Kauffmann}, {Khochfar}, {Kirkpatrick}, {Kocevski}, {Koekemoer}, {Labbe}, {Laporte}, {Larson}, {Lucas}, {Magee}, {Mason}, {McCracken}, {McLeod}, {McLure}, {Merlin}, {Mesinger}, {Milvang-Jensen}, {Newman}, {Oesch}, {Ouchi}, {Pacifici}, {Papovich}, {Peacock}, {Peeples}, {Pentericci}, {Perez-Gonzalez}, {Pirzkal}, {Pope}, {Pye}, {Reddy}, {Robertson}, {Salvato}, {Santini}, {Schaerer}, {Shapley}, {Simons}, {Smit}, {Smith}, {Snyder}, {Somerville}, {Stanway}, {Stefanon}, {Tasca}, {Tikkanen}, {Tresse}, {Trump}, {Whitaker},
  {Wilkins}, {Wright}, {Wyithe}, {van Dokkum}, \& {van der Werf}}]{2021jwst.prop.1837D}
{Dunlop}, J.~S., {Abraham}, R.~G., {Ashby}, M. L.~N., {et~al.} 2021, {PRIMER: Public Release IMaging for Extragalactic Research}, JWST Proposal. Cycle 1, ID. \#1837

\bibitem[{{Fazio} \& {Seds Team}(2011)}]{Fazio2011}
{Fazio}, G.~G., \& {Seds Team}. 2011, in Astronomical Society of the Pacific Conference Series, Vol. 446, Galaxy Evolution: Infrared to Millimeter Wavelength Perspective, ed. W.~{Wang}, J.~{Lu}, Z.~{Luo}, Z.~{Yang}, H.~{Hua}, \& Z.~{Chen}, 347

\bibitem[{{Finoguenov} {et~al.}(2007){Finoguenov}, {Guzzo}, {Hasinger}, {Scoville}, {Aussel}, {B{\"o}hringer}, {Brusa}, {Capak}, {Cappelluti}, {Comastri}, {Giodini}, {Griffiths}, {Impey}, {Koekemoer}, {Kneib}, {Leauthaud}, {Le F{\`e}vre}, {Lilly}, {Mainieri}, {Massey}, {McCracken}, {Mobasher}, {Murayama}, {Peacock}, {Sakelliou}, {Schinnerer}, {Silverman}, {Smol{\v{c}}i{\'c}}, {Taniguchi}, {Tasca}, {Taylor}, {Trump}, \& {Zamorani}}]{Finoguenov2007}
{Finoguenov}, A., {Guzzo}, L., {Hasinger}, G., {et~al.} 2007, \apjs, 172, 182, \dodoi{10.1086/516577}

\bibitem[{{Foreman-Mackey} {et~al.}(2013){Foreman-Mackey}, {Hogg}, {Lang}, \& {Goodman}}]{ForemanMackey2013}
{Foreman-Mackey}, D., {Hogg}, D.~W., {Lang}, D., \& {Goodman}, J. 2013, \pasp, 125, 306, \dodoi{10.1086/670067}

\bibitem[{{Giacconi} {et~al.}(1962){Giacconi}, {Gursky}, {Paolini}, \& {Rossi}}]{giacconi}
{Giacconi}, R., {Gursky}, H., {Paolini}, F.~R., \& {Rossi}, B.~B. 1962, \prl, 9, 439, \dodoi{10.1103/PhysRevLett.9.439}

\bibitem[{{Hale} {et~al.}(2018){Hale}, {Jarvis}, {Delvecchio}, {Hatfield}, {Novak}, {Smol{\v{c}}i{\'c}}, \& {Zamorani}}]{Hale2018}
{Hale}, C.~L., {Jarvis}, M.~J., {Delvecchio}, I., {et~al.} 2018, \mnras, 474, 4133, \dodoi{10.1093/mnras/stx2954}

\bibitem[{{Harikane} {et~al.}(2023){Harikane}, {Zhang}, {Nakajima}, {Ouchi}, {Isobe}, {Ono}, {Hatano}, {Xu}, \& {Umeda}}]{Harikane2023}
{Harikane}, Y., {Zhang}, Y., {Nakajima}, K., {et~al.} 2023, \apj, 959, 39, \dodoi{10.3847/1538-4357/ad029e}

\bibitem[{{Hasinger}(2020)}]{Hasinger2020}
{Hasinger}, G. 2020, \jcap, 2020, 022, \dodoi{10.1088/1475-7516/2020/07/022}

\bibitem[{{Helgason} {et~al.}(2014){Helgason}, {Cappelluti}, {Hasinger}, {Kashlinsky}, \& {Ricotti}}]{Helgason2014}
{Helgason}, K., {Cappelluti}, N., {Hasinger}, G., {Kashlinsky}, A., \& {Ricotti}, M. 2014, \apj, 785, 38, \dodoi{10.1088/0004-637X/785/1/38}

\bibitem[{{Helgason} {et~al.}(2012){Helgason}, {Ricotti}, \& {Kashlinsky}}]{helgason12}
{Helgason}, K., {Ricotti}, M., \& {Kashlinsky}, A. 2012, \apj, 752, 113, \dodoi{10.1088/0004-637X/752/2/113}

\bibitem[{{Hickox} \& {Markevitch}(2007)}]{hm06}
{Hickox}, R.~C., \& {Markevitch}, M. 2007, \apjl, 661, L117, \dodoi{10.1086/519003}

\bibitem[{{Hopkins} {et~al.}(2007){Hopkins}, {Richards}, \& {Hernquist}}]{Hopkins2007}
{Hopkins}, P.~F., {Richards}, G.~T., \& {Hernquist}, L. 2007, \apj, 654, 731, \dodoi{10.1086/509629}

\bibitem[{{Ilbert} {et~al.}(2006){Ilbert}, {Arnouts}, {McCracken}, {Bolzonella}, {Bertin}, {Le F{\`e}vre}, {Mellier}, {Zamorani}, {Pell{\`o}}, {Iovino}, {Tresse}, {Le Brun}, {Bottini}, {Garilli}, {Maccagni}, {Picat}, {Scaramella}, {Scodeggio}, {Vettolani}, {Zanichelli}, {Adami}, {Bardelli}, {Cappi}, {Charlot}, {Ciliegi}, {Contini}, {Cucciati}, {Foucaud}, {Franzetti}, {Gavignaud}, {Guzzo}, {Marano}, {Marinoni}, {Mazure}, {Meneux}, {Merighi}, {Paltani}, {Pollo}, {Pozzetti}, {Radovich}, {Zucca}, {Bondi}, {Bongiorno}, {Busarello}, {de La Torre}, {Gregorini}, {Lamareille}, {Mathez}, {Merluzzi}, {Ripepi}, {Rizzo}, \& {Vergani}}]{Ilbert2006}
{Ilbert}, O., {Arnouts}, S., {McCracken}, H.~J., {et~al.} 2006, \aap, 457, 841, \dodoi{10.1051/0004-6361:20065138}

\bibitem[{{Kashlinsky}(2016)}]{Kashlinsky2016}
{Kashlinsky}, A. 2016, \apjl, 823, L25, \dodoi{10.3847/2041-8205/823/2/L25}

\bibitem[{{Kashlinsky} {et~al.}(2012){Kashlinsky}, {Arendt}, {Ashby}, {Fazio}, {Mather}, \& {Moseley}}]{Kashlinsky2012}
{Kashlinsky}, A., {Arendt}, R.~G., {Ashby}, M.~L.~N., {et~al.} 2012, \apj, 753, 63, \dodoi{10.1088/0004-637X/753/1/63}

\bibitem[{Kashlinsky {et~al.}(2025)Kashlinsky, Arendt, Ashby, Kruk, \& Odegard}]{Kashlinsky2025}
Kashlinsky, A., Arendt, R.~G., Ashby, M. L.~N., Kruk, J., \& Odegard, N. 2025, The Astrophysical Journal Letters, 980, L12, \dodoi{10.3847/2041-8213/adad5e}

\bibitem[{Kashlinsky {et~al.}(2018)Kashlinsky, Arendt, Atrio-Barandela, Cappelluti, Ferrara, \& Hasinger}]{Kashlinsky2018}
Kashlinsky, A., Arendt, R.~G., Atrio-Barandela, F., {et~al.} 2018, Rev. Mod. Phys., 90, 025006, \dodoi{10.1103/RevModPhys.90.025006}

\bibitem[{{Kashlinsky} {et~al.}(2005){Kashlinsky}, {Arendt}, {Mather}, \& {Moseley}}]{Kashlinsky2005}
{Kashlinsky}, A., {Arendt}, R.~G., {Mather}, J., \& {Moseley}, S.~H. 2005, \nat, 438, 45, \dodoi{10.1038/nature04143}

\bibitem[{{Kocevski} {et~al.}(2023){Kocevski}, {Onoue}, {Inayoshi}, {Trump}, {Arrabal Haro}, {Grazian}, {Dickinson}, {Finkelstein}, {Kartaltepe}, {Hirschmann}, {Aird}, {Holwerda}, {Fujimoto}, {Juneau}, {Amor{\'\i}n}, {Backhaus}, {Bagley}, {Barro}, {Bell}, {Bisigello}, {Calabr{\`o}}, {Cleri}, {Cooper}, {Ding}, {Grogin}, {Ho}, {Hutchison}, {Inoue}, {Jiang}, {Jones}, {Koekemoer}, {Li}, {Li}, {McGrath}, {Molina}, {Papovich}, {P{\'e}rez-Gonz{\'a}lez}, {Pirzkal}, {Wilkins}, {Yang}, \& {Yung}}]{Kocevski2023}
{Kocevski}, D.~D., {Onoue}, M., {Inayoshi}, K., {et~al.} 2023, \apjl, 954, L4, \dodoi{10.3847/2041-8213/ace5a0}

\bibitem[{{Lambrides} {et~al.}(2024){Lambrides}, {Garofali}, {Larson}, {Ptak}, {Chiaberge}, {Long}, {Hutchison}, {Norman}, {McKinney}, {Akins}, {Berg}, {Chisholm}, {Civano}, {Cloonan}, {Endsley}, {Faisst}, {Gilli}, {Gillman}, {Hirschmann}, {Kartaltepe}, {Kocevski}, {Kokorev}, {Pacucci}, {Richardson}, {Stiavelli}, \& {Whalen}}]{Lambrides2024}
{Lambrides}, E., {Garofali}, K., {Larson}, R., {et~al.} 2024, arXiv e-prints, arXiv:2409.13047, \dodoi{10.48550/arXiv.2409.13047}

\bibitem[{{Li} {et~al.}(2018){Li}, {Cappelluti}, {Arendt}, {Hasinger}, {Kashlinsky}, \& {Helgason}}]{Li2018}
{Li}, Y., {Cappelluti}, N., {Arendt}, R.~G., {et~al.} 2018, \apj, 864, 141, \dodoi{10.3847/1538-4357/aad55a}

\bibitem[{{Limber}(1953)}]{Limber1953}
{Limber}, D.~N. 1953, \apj, 117, 134, \dodoi{10.1086/145672}

\bibitem[{{Matsumoto} {et~al.}(2011){Matsumoto}, {Seo}, {Jeong}, {Lee}, {Matsuura}, {Matsuhara}, {Oyabu}, {Pyo}, \& {Wada}}]{matsumoto}
{Matsumoto}, T., {Seo}, H.~J., {Jeong}, W.~S., {et~al.} 2011, \apj, 742, 124, \dodoi{10.1088/0004-637X/742/2/124}

\bibitem[{{Mitchell-Wynne} {et~al.}(2016){Mitchell-Wynne}, {Cooray}, {Xue}, {Luo}, {Brandt}, \& {Koekemoer}}]{mw16}
{Mitchell-Wynne}, K., {Cooray}, A., {Xue}, Y., {et~al.} 2016, \apj, 832, 104, \dodoi{10.3847/0004-637X/832/2/104}

\bibitem[{{Moretti} {et~al.}(2012){Moretti}, {Vattakunnel}, {Tozzi}, {Salvaterra}, {Severgnini}, {Fugazza}, {Haardt}, \& {Gilli}}]{moretti}
{Moretti}, A., {Vattakunnel}, S., {Tozzi}, P., {et~al.} 2012, \aap, 548, A87, \dodoi{10.1051/0004-6361/201219921}

\bibitem[{{Natarajan} {et~al.}(2024){Natarajan}, {Pacucci}, {Ricarte}, {Bogd{\'a}n}, {Goulding}, \& {Cappelluti}}]{Natarajan2024}
{Natarajan}, P., {Pacucci}, F., {Ricarte}, A., {et~al.} 2024, \apjl, 960, L1, \dodoi{10.3847/2041-8213/ad0e76}

\bibitem[{{Pacucci} \& {Narayan}(2024)}]{Pacucci2024}
{Pacucci}, F., \& {Narayan}, R. 2024, \apj, 976, 96, \dodoi{10.3847/1538-4357/ad84f7}

\bibitem[{{Pacucci} {et~al.}(2023){Pacucci}, {Nguyen}, {Carniani}, {Maiolino}, \& {Fan}}]{Pacucci2023}
{Pacucci}, F., {Nguyen}, B., {Carniani}, S., {Maiolino}, R., \& {Fan}, X. 2023, \apjl, 957, L3, \dodoi{10.3847/2041-8213/ad0158}

\bibitem[{{Paquereau} {et~al.}(2025){Paquereau}, {Laigle}, {McCracken}, {Shuntov}, {Ilbert}, {Akins}, {Allen}, {Arango- Togo}, {Berman}, {Bethermin}, {Casey}, {McCleary}, {Dubois}, {Drakos}, {Faisst}, {Franco}, {Harish}, {Jespersen}, {Kartaltepe}, {Koekemoer}, {Kokorev}, {Lambrides}, {Larson}, {Liu}, {Le Borgne}, {Lewis}, {McKinney}, {Mercier}, {Rhodes}, {Robertson}, {Toft}, {Trebitsch}, {Tresse}, \& {Weaver}}]{Paquereau2025}
{Paquereau}, L., {Laigle}, C., {McCracken}, H.~J., {et~al.} 2025, arXiv e-prints, arXiv:2501.11674.
\newblock \doarXiv{2501.11674}

\bibitem[{{Powell} {et~al.}(2020){Powell}, {Urry}, {Cappelluti}, {Johnson}, {LaMassa}, {Ananna}, \& {Kollmann}}]{Powell2020}
{Powell}, M.~C., {Urry}, C.~M., {Cappelluti}, N., {et~al.} 2020, \apj, 891, 41, \dodoi{10.3847/1538-4357/ab6e65}

\bibitem[{{Reynolds} {et~al.}(2023){Reynolds}, {Kara}, {Mushotzky}, {Ptak}, {Koss}, {Williams}, {Allen}, {Bauer}, {Bautz}, {Bogadhee}, {Burdge}, {Cappelluti}, {Cenko}, {Chartas}, {Chan}, {Corrales}, {Daylan}, {Falcone}, {Foord}, {Grant}, {Habouzit}, {Haggard}, {Herrmann}, {Hodges-Kluck}, {Kargaltsev}, {King}, {Kounkel}, {Lopez}, {Marchesi}, {McDonald}, {Meyer}, {Miller}, {Nynka}, {Okajima}, {Pacucci}, {Russell}, {Safi-Harb}, {Strassun}, {Trindade Falc{\~a}o}, {Walker}, {Wilms}, {Yukita}, \& {Zhang}}]{axis}
{Reynolds}, C.~S., {Kara}, E.~A., {Mushotzky}, R.~F., {et~al.} 2023, in Society of Photo-Optical Instrumentation Engineers (SPIE) Conference Series, Vol. 12678, UV, X-Ray, and Gamma-Ray Space Instrumentation for Astronomy XXIII, ed. O.~H. {Siegmund} \& K.~{Hoadley}, 126781E, \dodoi{10.1117/12.2677468}

\bibitem[{{Ricarte} \& {Natarajan}(2018)}]{ricarte18}
{Ricarte}, A., \& {Natarajan}, P. 2018, \mnras, 474, 1995, \dodoi{10.1093/mnras/stx2851}

\bibitem[{{Ricarte} {et~al.}(2019){Ricarte}, {Pacucci}, {Cappelluti}, \& {Natarajan}}]{Ricarte2019}
{Ricarte}, A., {Pacucci}, F., {Cappelluti}, N., \& {Natarajan}, P. 2019, \mnras, 489, 1006, \dodoi{10.1093/mnras/stz1891}

\bibitem[{{Salvaterra} {et~al.}(2012){Salvaterra}, {Haardt}, {Volonteri}, \& {Moretti}}]{Salvaterra2012}
{Salvaterra}, R., {Haardt}, F., {Volonteri}, M., \& {Moretti}, A. 2012, \aap, 545, L6, \dodoi{10.1051/0004-6361/201219965}

\bibitem[{{Sersic}(1968)}]{Sersic1968}
{Sersic}, J.~L. 1968, {Atlas de Galaxias Australes}

\bibitem[{{Shankar} {et~al.}(2009){Shankar}, {Weinberg}, \& {Miralda-Escud{\'e}}}]{Shankar2009}
{Shankar}, F., {Weinberg}, D.~H., \& {Miralda-Escud{\'e}}, J. 2009, \apj, 690, 20, \dodoi{10.1088/0004-637X/690/1/20}

\bibitem[{{Sheth} {et~al.}(2001){Sheth}, {Mo}, \& {Tormen}}]{SMT2001}
{Sheth}, R.~K., {Mo}, H.~J., \& {Tormen}, G. 2001, \mnras, 323, 1, \dodoi{10.1046/j.1365-8711.2001.04006.x}

\bibitem[{{Shuntov} {et~al.}(2024){Shuntov}, {Ilbert}, {Toft}, {Arango-Toro}, {Akins}, {Casey}, {Franco}, {Harish}, {Kartaltepe}, {Koekemoer}, {McCracken}, {Paquereau}, {Laigle}, {Bethermin}, {Dubois}, {Drakos}, {Faisst}, {Gozaliasl}, {Gillman}, {Hayward}, {Hirschmann}, {Huertas-Company}, {Jespersen}, {Jin}, {Kokorev}, {Lambrides}, {Le Borgne}, {Liu}, {Magdis}, {Massey}, {McPartland}, {Mercier}, {McCleary}, {McKinney}, {Oesch}, {Rhodes}, {Rich}, {Robertson}, {Sanders}, {Trebitsch}, {Tresse}, {Valentino}, {Vijayan}, {Weaver}, {Weibel}, \& {Wilkins}}]{Shuntov2024}
{Shuntov}, M., {Ilbert}, O., {Toft}, S., {et~al.} 2024, arXiv e-prints, arXiv:2410.08290, \dodoi{10.48550/arXiv.2410.08290}

\bibitem[{{Soltan}(1982)}]{Soltan1982}
{Soltan}, A. 1982, \mnras, 200, 115, \dodoi{10.1093/mnras/200.1.115}

\bibitem[{{Yue} {et~al.}(2013){Yue}, {Ferrara}, {Salvaterra}, \& {Chen}}]{Yue2013}
{Yue}, B., {Ferrara}, A., {Salvaterra}, R., \& {Chen}, X. 2013, \mnras, 431, 383, \dodoi{10.1093/mnras/stt174}

\end{thebibliography}
\bibliographystyle{aasjournal}



\end{document}